\documentclass[%
 aip,
 amsmath,amssymb,
 reprint,%
floatfix]{revtex4-1}

\usepackage{graphicx}
\usepackage{dcolumn}
\usepackage{bm}

\usepackage[utf8]{inputenc}
\usepackage[T1]{fontenc}
\usepackage{mathptmx}
\usepackage{etoolbox}
\usepackage{color}

\newcommand{\COMMENTED}[1]{}

\makeatletter
\def\@email#1#2{%
 \endgroup
 \patchcmd{\titleblock@produce}
  {\frontmatter@RRAPformat}
  {\frontmatter@RRAPformat{\produce@RRAP{*#1\href{mailto:#2}{#2}}}\frontmatter@RRAPformat}
  {}{}
}%
\makeatother
\begin{document}

\preprint{AIP/123-QED}

\title{Monte Carlo Methods in the Manifold of Hartree-Fock-Bogoliubov Wave Functions}
\author{Ettore Vitali}
 \affiliation{Department of Physics, California State University Fresno, Fresno, California 93720, USA}
 \email{evitali@mail.fresnostate.edu}
\author{Peter Rosenberg}%

\author{Shiwei Zhang}
\affiliation{%
Center for Computational Quantum Physics, Flatiron Institute, 
New York, NY10010, USA
}%

\date{\today}

\begin{abstract}
We explore the possibility to implement random walks in the manifold of  Hartree-Fock-Bogoliubov wave functions. 
The goal is to extend state-of-the-art 
quantum Monte Carlo approaches, 
in particular the constrained-path auxiliary-field quantum Monte Carlo technique, to systems where finite pairing order parameters or complex pairing mechanisms, e.g., Fulde-Ferrell-Larkin-Ovchinnikov (FFLO) pairing or triplet pairing, may be expected. Leveraging the flexibility  to define a vacuum state tailored to the physical problem,
we discuss a method 
to use imaginary-time evolution of Hartree-Fock-Bogoliubov states to compute ground state correlations, 
extending beyond 
situations spanned by 
current formalisms.
Illustrative examples are provided.
\end{abstract}

\maketitle

%

\section{\label{sec:introduction} Introduction}

Strongly correlated many-body systems underlie a wide variety of exotic physical phenomena, including spin liquids, 
high-temperature 
superconductivity, and other unconventional pairing states, among other examples. These systems 
pose a significant theoretical and 
computational challenge, due primarily to the high-dimensional nature of the problem and the lack of reliable approximative techniques. 
To gain a better understanding of these systems and their emergent behavior therefore requires the continual development 
of robust, high-accuracy numerical tools and methodologies. One cutting-edge method with proven capabilities in the treatment of strongly 
correlated many-body systems is auxiliary-field quantum Monte Carlo (AFQMC) [\onlinecite{PhysRevB.55.7464}]. The AFQMC technique has been applied
to a broad range of physical systems and problems, from lattice models in the context of condensed matter physics [\onlinecite{PhysRevB.78.165101},\onlinecite{doi:10.1126/science.adh7691}], to ab-initio quantum chemistry 
calculations [\onlinecite{PhysRevLett.90.136401},\onlinecite{https://doi.org/10.1002/wcms.1364},\onlinecite{PhysRevX.10.031058}], to fermionic cold atoms [\onlinecite{PhysRevA.92.033603}]. In some cases 
the method yields numerically exact results,
while in other cases it provides an excellent balance between high accuracy and the capability to reach large sizes for realistic systems [\onlinecite{PhysRevX.5.041041}].

The AFQMC algorithm carries out a random walk in the manifold of independent-fermion wave functions, with the walkers most commonly in the form of Slater determinants. 
This formulation has provided accurate results for various models [\onlinecite{PhysRevLett.117.040401}, \onlinecite{PhysRevLett.119.265301}, \onlinecite{PhysRevLett.128.203201}], chemical systems [\onlinecite{PhysRevX.7.031059}], and real materials [\onlinecite{PhysRevB.75.245123},\onlinecite{PhysRevLett.114.226401}]. However, for many systems, it is desirable to have a more flexible form for the walkers 
in order to capture the physics of the system more reliably and accurately. In the context of systems supporting exotic pairing states, a natural extension is to generalize from Slater determinants to Hartree-Fock-Bogoliubov (HFB) wave functions, which provides a larger manifold for the random walk. This generalization enables the sampling of wave functions that contain information about fermionic pairing, which is essential to achieving more accurate results for systems where pairing is expected. 
Importantly, the use of HFB wave functions allows the particle number to fluctuate, which permits the direct computation of pairing order parameters. This is a crucial advantage over simulations with fixed particle number that must instead rely on correlation functions, which typically have small signals  [\onlinecite{doi:10.1126/science.adh7691}, \onlinecite{PhysRevX.10.031016}], to detect orders.

While some work exists in this direction, each of the previously developed approaches has limitations that curtail its applicability. For example, the method devised in
[\onlinecite{PhysRevA.100.023621}], that provided robust evidence of a Fulde-Ferrell-Larkin-Ovchinnikov (FFLO) state in spin-polarized optical lattice systems [\onlinecite{PhysRevLett.128.203201}], is limited to the case of singlet pairing and $U(1)$ symmetry, which prevents the direct 
computation of order parameters. On the other hand, pairing order parameters have been computed directly, relying on a partial particle-hole transformation [\onlinecite{PhysRevX.10.031016}], but such a transformation was specifically tailored to the $d$-wave problem studied in [\onlinecite{PhysRevX.10.031016}]. 
The
possibility of working directly with HFB wave functions is explored in [\onlinecite{PhysRevB.95.045144}], but the main focus and all the applications were on 
fully paired states.
In the work of Juillet et al. [\onlinecite{Juillet_2017}], spin-polarized systems are considered within a similar formalism, still relying on HFB wave functions. Their work focused however on the study of $d$-wave pairing in cuprates through a partial particle-hole transformation, and the results 
were primarily
energy calculations within a projection to a state with fixed spin-polarization.

In this work we generalize the approaches pioneered in [\onlinecite{PhysRevB.95.045144}] and [\onlinecite{Juillet_2017}]. Taking advantage of the freedom to define different 
vacuum states, we design an algorithm to study imaginary time dynamics in the manifold of HFB wave functions by random walks based on rotations from 
a special vacuum tailored to the hamiltonian of the system.
We demonstrate that the algorithm preserves
polynomial scaling analogous to state-of-the-art AFQMC
while exploring a larger manifold. We provide the details of such an extension, test the methodology to address possible numerical stabilization issues, and show a comparison with exact diagonalization. 

The paper is organized as follows. 
In Sec.~\ref{sec:mathematics} we 
briefly outline some 
mathematical background.
Then in Sec.~\ref{sec:formalism}
we present the mathematical formalism for our approach including all necessary theorems. Additional proofs are provided in the Appendices. 
In Sec.~\ref{sec:hfbmotion} we outline in concrete terms  
a numerically stable method to perform imaginary time evolution driven by quadratic hamiltonians in the manifold of HFB wave functions. Then, in Sec.~\ref{sec:qmcmotion} we discuss the Monte Carlo sampling required to treat correlated hamiltonians and show a comparison with exact diagonalization. Finally, in Sec.~\ref{sec:conclusion}, we conclude and discuss potential applications.

\section{\label{sec:mathematics} The Mathematical Background}
\subsection{\label{sec:notations} Notations and Conventions} 
We build our mathematical description of a collection of fermions starting from a single-particle complex Hilbert space, $\mathcal{H}$, with dimension $M$. Throughout this work, we will assume $M<+\infty$, which can always be achieved through suitable regularization techniques, for example focusing on fermions moving on a finite lattice or a single-particle basis set.
We define an orthonormal basis $\{ | \alpha \rangle \}_{\alpha=1,\dots,M} \subset \mathcal{H}$ and construct the corresponding algebra of $2M$ creation and destruction operators $\{ \hat{c}^{\dagger}_{\alpha} , \hat{c}^{}_{\alpha} \}_{\alpha=1,\dots,M}$, satisfying the canonical anticommutation relations.
The many-body theory is formulated in the fermionic Fock space:
\begin{equation}
    \label{mathematics:FockSpace}
    \mathcal{F}\left(\mathcal{H}\right) = \mathcal{H}^{(0)} \oplus \mathcal{H}^{(1)} \oplus \dots \oplus \mathcal{H}^{(N)} \oplus \dots \oplus \mathcal{H}^{(M)}
\end{equation}
where $\mathcal{H}^{(0)}$ is a one-dimensional space generated by a vacuum state $|0\rangle$, annihilated by all the destruction operators $\hat{c}^{}_{\alpha} |0\rangle = 0$, while $\mathcal{H}^{(N)}$ is the ${M\choose N}$-dimensional anti-symmetric $N$-tensor product space, made of states with a well defined number of particles, $N$.  

\subsection{\label{sec:hfbmanifold} The manifold of Hartree-Fock-Bogoliubov Wave Functions}
In this work we focus on the manifold of 
HFB wave functions, a subset of the Fock space of the system, which we denote as $\mathcal{M}_{HFB}$. By definition, $| \omega \rangle \in \mathcal{M}_{HFB}$ if there exists a unitary transformation, commonly referred to as a Bogoliubov transformation, of the form:
\begin{equation}
\label{hfbmanifold:HFB}
\left( \hat{\gamma}^{\dagger} \,\, \hat{\gamma}^{} \right)
= \, \left( \hat{c}^{\dagger} \,\, \hat{c}^{} \right) \,\,
\left(
\begin{array}{cc}
 U & V^{\star} \\
 V & U^{\star}
\end{array}
\right)
\end{equation}
such that $| \omega \rangle$ is the vacuum of the new family of operators, that is, if $\hat{\gamma}^{}_{\alpha} | \omega \rangle = 0$.
In \eqref{hfbmanifold:HFB}, $U$ and $V$ are $M\times M$ complex matrices, satisfying the following conditions that guarantee that the transformation in \eqref{hfbmanifold:HFB} is unitary:
\begin{equation}
\label{hfbmanifold:UVprop}
\begin{split}
& U U^{\dagger} + V^{\star}V^{T} = \mathbb{I}, \quad U V^{\dagger} + V^{\star}U^{T} = 0 \\
& U^{\dagger} U + V^{\dagger}V = \mathbb{I}, \quad U^{\dagger} V^{\star} + V^{\dagger}U^{\star} = 0 \\
\end{split}
\end{equation}
In the above, $\mathbb{I}$ is the $M \times M$ identity matrix.

A special example of an HFB wave function is the vacuum $|0\rangle$ itself, the state with zero fermions, corresponding to $V=0$ and any choice of unitary matrix $U$.
In addition, any $N$-particle Slater determinant of the form $| \phi \rangle = \hat{\beta}^{\dagger}_{N} \dots \hat{\beta}^{\dagger}_1 | 0 \rangle$, with $\hat{\beta}^{\dagger}_{i} = \sum_{\alpha} \phi_{i}(\alpha) \hat{c}^{\dagger}_{\alpha}$, belongs to $\mathcal{M}_{HFB}$, as can be seen easily by defining the operators:
\begin{equation}
\hat{\gamma}_i = 
\begin{cases}
\hat{\beta}^{\dagger}_{i}, \quad i = 1, \dots, N \\
\hat{\beta}^{}_{i}, \quad i = N+1, \dots, M
\end{cases}
\end{equation}
where, for $i = N+1, \dots, M$, the operators $\hat{\beta}^{}_{i} = \sum_{\alpha} \phi_{i}^{\star}(\alpha) \hat{c}^{}_{\alpha}$ are destruction operators related to 
orbitals $\phi_i$ that are orthogonal to the occupied ones in $| \phi \rangle$. Another very important subset of $\mathcal{M}_{HFB}$ is the set of fully paired HFB wave functions, which can be written as:
\begin{equation}
\label{hfbmanifold:fullypaired}
| \omega_{\rm paired} \rangle = \mathcal{C} \exp\left( \sum_{\alpha < \beta} Z_{\alpha,\beta} \hat{c}^{\dagger}_{\alpha} \hat{c}^{\dagger}_{\beta} \right) \, | 0 \rangle
\end{equation}
where $\mathcal{C}$ is a normalization constant, while $Z$ is an anti-symmetric $M \times M$ matrix and $| 0 \rangle$ is the vacuum state. Such wave functions have non-zero components in $\mathcal{H}^{(N)}$ if and only if $N$ is even, and describe collections of Cooper pairs, generalizing the well-known BCS wave function. 
The relation between the matrix $Z$ in \eqref{hfbmanifold:fullypaired} and the matrices $U,V$ corresponding to $| \omega_{\rm paired} \rangle$ is $Z = \left( V U^{-1} \right)^{\star}$. Note that, if $\det(U)=0$, then the HFB wave function cannot be expressed in the form given in \eqref{hfbmanifold:fullypaired}.
We find it useful to mention the important class of particle-number projected wave functions, which is often referred to in quantum chemistry as anti-symmetrized geminal power (AGP) [\onlinecite{Coleman1965},\onlinecite{rin80}]:
\begin{equation}
\label{hfbmanifold:AGP}
| \omega_{\rm AGP} \rangle = \mathcal{C'} 
( \sum_{\alpha < \beta} Z_{\alpha,\beta} \hat{c}^{\dagger}_{\alpha} \hat{c}^{\dagger}_{\beta}
)^{N/2} \, | 0 \rangle\,,
\end{equation}
Such wave functions can be obtained from HFB states by simply projecting them onto $\mathcal{H}^{(N)}$.

\section{\label{sec:formalism} Formalism for mean-field and correlated calculations}

The family of wave functions \eqref{hfbmanifold:fullypaired} does not include HFB wave functions with unpaired fermions, which 
are important to include in order
to study spin-polarized systems. 
We observe that 
it is possible to directly generalize \eqref{hfbmanifold:fullypaired} by replacing $|0 \rangle$ with an arbitrary HFB wave function, say $|\omega_{v} \rangle$, playing the role of a ``new vacuum'' of the theory (hence the label ``$v$''), and corresponding to a particular 
Bogoliubov transformation matrix:
\begin{equation}
\left( \hat{\gamma}^{\dagger} \,\, \hat{\gamma}^{} \right)
= \, \left( \hat{c}^{\dagger} \,\, \hat{c}^{} \right) \,\,
\left(
\begin{array}{cc}
 U_v & V_v^{\star} \\
 V_v & U_v^{\star}
\end{array}
\right)
\end{equation}
Practically, for example, we can always 
construct $| \omega_{v} \rangle$ by performing an HFB calculation tailored to the physical problem we are studying.

Our investigation will focus on the subset $\mathcal{S}_v \subset \mathcal{M}_{HFB}$ of states that are non-orthogonal to our new vacuum:
\begin{equation}
\mathcal{S}_v = \left\{ | \omega \rangle \in \mathcal{M}_{HFB} \, : \, \langle \omega_v \, | \, \omega \rangle \neq 0 \right\}
\end{equation}
Given any HFB wave function $| \omega \rangle$, related to an HFB transformation defined by matrices $U$ and $V$,
we can 
check whether $| \omega \rangle \in \mathcal{S}_v$ by using Onishi's formula [\onlinecite{rin80}], which implies that $| \omega \rangle \in \mathcal{S}_v$ if and only if $\det\left(  {U}^{\dagger}_v U +  {V}_v^{\dagger} V \right) \neq 0$.
Every wave function $| \omega \rangle \in \mathcal{S}_v$ can then be expressed in the following form, based on Thouless' theorem [\onlinecite{rin80}] and generalizing \eqref{hfbmanifold:fullypaired}:
\COMMENTED{
\begin{equation}
\label{hfbmanifold:thoulessgeneral}
\begin{split}
& | \omega \rangle = \langle \omega_v \, | \, \omega \rangle \, \exp\left( \hat{Z} \right) \, | {\omega}_v \rangle, 
\quad \hat{Z} = \sum_{\alpha < \beta} Z_{\alpha,\beta} \, \hat{\gamma}^{\dagger}_\alpha \hat{\gamma}^{\dagger}_\beta \\
& Z = \left( \left(  {V}_v^{T}  U +  {U}_v^{T} V \right) \left( {U}^{\dagger}_v U +  {V}_v^{\dagger} V  \right)^{-1}\right)^{\star}
\end{split}
\end{equation}
}
\begin{equation}
\label{hfbmanifold:thoulessgeneral}
 | \omega \rangle = \langle \omega_v \, | \, \omega \rangle \, \exp\left( \hat{Z} \right) \, | {\omega}_v \rangle\,, 
 \end{equation}
 where $\hat{Z} = \sum_{\alpha < \beta} Z_{\alpha,\beta} \, \hat{\gamma}^{\dagger}_\alpha \hat{\gamma}^{\dagger}_\beta$ with the
 $M \times M$ complex matrix $Z$ defined as:
\begin{equation}
\label{hfbmanifold:thoulessgeneral-Z-def} 
 Z = \left( \left(  {V}_v^{T}  U +  {U}_v^{T} V \right) \left( {U}^{\dagger}_v U +  {V}_v^{\dagger} V  \right)^{-1}\right)^{\star}\,,
\end{equation}
which is anti-symmetric.
The above expression provides a method to construct $| \omega \rangle$ by exciting quasiparticles 
with respect to 
$| \omega_{v} \rangle$.
We observe that the matrices:
\begin{equation}
\label{mathbbUV}
\mathbb{V} = {V}_v^{T}  U +  {U}_v^{T} V , \quad \mathbb{U} = {U}^{\dagger}_v U +  {V}_v^{\dagger} V 
\end{equation}
build the transformation between the Bogoliubov operators defining $| \omega_{v} \rangle$ and those defining $| \omega \rangle$, and thus play a role similar to $(U,V)$ 
when the vacuum is $|0\rangle$ (instead of $| \omega_{v} \rangle$).
For this reason, they must also satisfy the conditions given in \eqref{hfbmanifold:UVprop}.

There are a few natural parametrizations of $\mathcal{S}_v$: the Bogoliubov matrices $(U,V)$, or the corresponding $(\mathbb{U},\mathbb{V})$ satisfying the constraints \eqref{hfbmanifold:UVprop}, and the anti-symmetic matrix $Z$ in \eqref{hfbmanifold:thoulessgeneral}:
\begin{equation}
\label{parametrize:Z}
|\omega\rangle \in \mathcal{S}_v \subset  \mathcal{M}_{HFB} \longleftrightarrow (U,V)\longleftrightarrow  (\mathbb{U},\mathbb{V}) \longleftrightarrow Z
\end{equation}
The ability to transform between these parametrizations is of convenience in the framework we present below.
In particular, we observe that the choice $Z=0$ corresponds to $|\omega_v \rangle$ itself, suggesting an interpretation of $|\omega_v \rangle$ as the ``origin'' of a ``coordinate system''. This matrix $Z$ turns out to be very helpful for calculations, as it permits the straightforward computation of overlaps and density matrices. The computation of overlaps and density matrices is a key component of the method we develop here. We discuss the details of these calculations in the following sections. 

\subsection{Overlaps}

Consider $| \omega_0 \rangle \in \mathcal{S}_v$ and $| \omega_1 \rangle \in \mathcal{S}_v$.
We can write:
\begin{equation}
\begin{split}
& | \omega_l \rangle = \langle \omega_v \, | \, \omega_l \rangle \, e^{\hat{Z}^{(l)}} \, | {\omega}_v \rangle, 
\quad \hat{Z}^{(l)} = \sum_{\alpha < \beta} Z^{(l)}_{\alpha,\beta} \, \hat{\gamma}^{\dagger}_\alpha \hat{\gamma}^{\dagger}_\beta \\
& Z^{(l)} = \left( \left(  {V}_v^{T}  U_l +  {U}_v^{T} V_l \right) \left( {U}^{\dagger}_v U_l +  {V}_v^{\dagger} V_l  \right)^{-1}\right)^{\star}\,,
\end{split}
\end{equation}
with $l=0$ or $1$.
The following identity can be proved [\onlinecite{Porro2022}]:
\begin{equation}
\begin{split}
& \langle \omega_0 \, | \, \omega_1 \rangle
= \langle \omega_0 \, | \, \omega_v \rangle \, \langle \omega_v \, | \, \omega_1 \rangle
\,  \langle  \omega_v \, | e^{\hat{Z}^{(0) \, \dagger}} \, e^{\hat{\tilde{Z}}^{(1)}} \, | \,  \omega_v \rangle  \\
& = \langle \omega_0 \, | \, \omega_v \rangle \, \langle \omega_v \, | \, \omega_1 \rangle \, (-1)^{M(M+1)/2} \, 
 pf \left(
\begin{array}{cc}
{Z}^{(1)} & - \mathbb{I}\\
\mathbb{I} &   - Z^{(0) \, \star}
\end{array}
\right)
\end{split}
\label{eq:ovlp_normalized}
\end{equation}
To avoid computing the normalization factors contained in \eqref{eq:ovlp_normalized}, we instead consider the unnormalized states:
\begin{equation}
| \phi_l \rangle = e^{\hat{Z}^{(l)}} \, | {\omega}_v \rangle
= \exp\left(  \sum_{\alpha < \beta} Z^{(l)}_{\alpha,\beta} \, \hat{\gamma}^{\dagger}_\alpha \hat{\gamma}^{\dagger}_\beta \right) \, | {\omega}_v \rangle
\end{equation}
The formula for the overlap is then:
\begin{equation}
\label{overlap:pfaffian}
\langle \phi_0 \, | \, \phi_1 \rangle = (-1)^{M(M+1)/2} \, 
 pf \left(
\begin{array}{cc}
{Z}^{(1)} & - \mathbb{I}\\
\mathbb{I} &   - Z^{(0) \, \star}
\end{array}
\right)
\end{equation}
Note that, by construction, $\langle \omega_v | \phi_l \rangle = 1$ for $l = 0, 1$.

\subsection{Density matrices}

We now introduce the following definitions, relying on the family of operators $\{ \hat{\gamma}_{k}, \hat{\gamma}^{\dagger}_{k} \}_{k=1,\dots,M}$ related to the vacuum $| \omega_{v} \rangle$, i.e., $\hat{\gamma}_{k} \, | \omega_{v} \rangle = 0$.
We let:
\begin{equation}
\begin{split}
& \rho_{k_1 k_2} \buildrel{def}\over{=} \frac{\langle \phi_0 \, | \hat{\gamma}^{\dagger}_{k_2} \hat{\gamma}^{}_{k_1} \, | \, \phi_1 \rangle }{\langle \phi_0 \, | \, \phi_1 \rangle }, \quad
\kappa_{k_1 k_2} \buildrel{def}\over{=} \frac{\langle \phi_0 \, | \hat{\gamma}^{}_{k_2} \hat{\gamma}^{}_{k_1} \, | \, \phi_1 \rangle }{\langle \phi_0 \, | \, \phi_1 \rangle } \\
& \overline{\kappa}^{\star}_{k_1 k_2} \buildrel{def}\over{=} - \frac{\langle \phi_0 \, | \hat{\gamma}^{\dagger}_{k_2} \hat{\gamma}^{\dagger}_{k_1} \, | \, \phi_1 \rangle }{\langle \phi_0 \, | \, \phi_1 \rangle }, \quad
\sigma^{\star}_{k_1 k_2} \buildrel{def}\over{=} -\frac{\langle \phi_0 \, | \hat{\gamma}^{}_{k_2} \hat{\gamma}^{\dagger}_{k_1} \, | \, \phi_1 \rangle }{\langle \phi_0 \, | \, \phi_1 \rangle }
\end{split}
\end{equation}
The following results have been proved in [\onlinecite{Porro2022}]:
\begin{align}
\rho_{k_1 k_2} &= - \left( Z^{(1)} \, \left( \mathbb{I} - Z^{(0) \, \star} Z^{(1)} \right)^{-1} \, Z^{(0) \, \star} \right)_{k_1, k_2} \label{rhok} \\
\kappa_{k_1 k_2}  &=  \left( Z^{(1)} \, \left( \mathbb{I} - Z^{(0) \, \star} Z^{(1)} \right)^{-1} \,  \right)_{k_1, k_2} \\
\overline{\kappa}^{\star}_{k_1 k_2}   &=  \left(  \left( \mathbb{I} - Z^{(0) \, \star} Z^{(1)} \right)^{-1} \,  Z^{(0) \, \star}  \right)_{k_1, k_2} \\
\sigma^{\star}_{k_1 k_2}   &=  - \left( \mathbb{I} - Z^{(0) \, \star} Z^{(1)} \right)^{-1}_{k_1, k_2} \label{sigmak}
\end{align}
From the above density matrices, it is straightforward to transform back to the original set of creation and destruction operators $\hat{c}^{\dagger}$ and $\hat{c}$.
For example, the matrix element of the one-body density matrix becomes:
\begin{equation}
\label{dmattransformback}
\begin{split}
& G_{\alpha \beta} = \frac{\langle \phi_0 \, | \hat{c}^{\dagger}_{\beta} \, \hat{c}^{}_{\alpha} \, | \, \phi_1 \rangle }{\langle \phi_0 \, | \, \phi_1 \rangle } \\
& = \sum_{l s}  \left(U_v^{\star}\right)_{\beta l} \left(V_v^{\star}\right)_{\alpha s} \, \left( - \overline{\kappa}^{\star}_{s l}  \right)
+ \sum_{l s}  \left(U_v^{\star}\right)_{\beta l}  \left(U_v^{}\right)_{\alpha s} \, \rho_{sl} \\
& + \sum_{l s} \left(V_v^{}\right)_{\beta l} \left(V_v^{\star}\right)_{\alpha s}  \, \left( - {\sigma}^{\star}_{s l}  \right)
+ \sum_{l s}  \left(V_v^{}\right)_{\beta l}  \left(U_v^{}\right)_{\alpha s} \, \kappa_{s l}
\end{split}
\end{equation}
For the pairing tensor we get:
\begin{equation}
\label{pairingtransformback}
\begin{split}
& \tau_{\alpha \beta} = \frac{\langle \phi_0 \, | \hat{c}^{}_{\beta} \, \hat{c}^{}_{\alpha} \, | \, \phi_1 \rangle }{\langle \phi_0 \, | \, \phi_1 \rangle } \\
& = \sum_{l s}  \left(V_v^{\star}\right)_{\beta l} \left(V_v^{\star}\right)_{\alpha s} \, \left( - \overline{\kappa}^{\star}_{s l}  \right)
+ \sum_{l s}  \left(V_v^{\star}\right)_{\beta l}  \left(U_v^{}\right)_{\alpha s} \, \rho_{sl} \\
& + \sum_{l s} \left(U_v^{}\right)_{\beta l} \left(V_v^{\star}\right)_{\alpha s}  \, \left( - {\sigma}^{\star}_{s l}  \right)
+ \sum_{l s}  \left(U_v^{}\right)_{\beta l}  \left(U_v^{}\right)_{\alpha s} \, \kappa_{s l}
\end{split}
\end{equation}
Note that the above transformation formulas, which allow us to move from the algebra of the new vacuum to the original algebra of creation and destruction operators (say in the basis of position and spin), depend only on the matrices $U_v$ and $V_v$. These matrices are known and stored at the very beginning of the calculation, while the states $|\phi_l\rangle$ enter only through the density matrices.

\subsection{\label{sec:hfbexpo} Exponential of quadratic operators in the manifold of HFB wave functions}

Another essential component of the approach we develop in this work is the computation of exponentials of quadratic operators.
In this section we provide the details of the construction of such operators for HFB wave functions.

Let us consider a non-normalized state in $\mathcal{S}_v$:
\begin{equation}
| \phi \rangle = e^{\hat{Z} }\, | {\omega}_v \rangle
= \exp\left(  \sum_{\alpha < \beta} Z_{\alpha,\beta} \, \hat{\gamma}^{\dagger}_\alpha \hat{\gamma}^{\dagger}_\beta \right) \, | {\omega}_v \rangle\,.
\end{equation}
We would like to study the map:
\begin{equation}
\label{exponential:map}
| \phi \rangle \in \mathcal{M}_{HFB} \rightarrow | \phi^{\prime} \rangle  \buildrel{def}\over{=}
\exp\left( \hat{O} \right) \, | \phi \rangle\,,
\end{equation}
where, following Ref.~[\onlinecite{PhysRevB.95.045144}], we take $\hat{O}$ to be a general ``quadratic'' operator of the form:
\begin{equation}
\label{hfbexpo:onebprop}
\hat{O} = \sum_{\alpha,\beta=1}^{M} t_{\alpha \beta} \, \hat{c}^{\dagger}_{\alpha} \hat{c}^{}_{\beta} +  \sum_{\alpha > \beta} {\Delta}_{\alpha \beta} \, \hat{c}^{\dagger}_{\alpha} \hat{c}^{\dagger}_{\beta} + \sum_{\alpha > \beta} \tilde{\Delta}_{\alpha \beta} \, \hat{c}^{}_{\alpha} \hat{c}^{}_{\beta}\,.  
\end{equation}
The matrix elements 
in the $M \times M$ complex matrices 
satisfy 
$\Delta^{T} = - \Delta$ and $\tilde{\Delta}^{T} = - \tilde{\Delta}$ and,
for 
hermitian operators, 
$t = t^{\dagger}$ and $\tilde{\Delta} = - \Delta^{\star}$.

We know that $| \phi \rangle$ is a vacuum of a family of operators:
\begin{equation}
 \hat{\beta}^{}_l \, | \phi \rangle = 0, \quad \left( \hat{\beta}^{\dagger} \,\, \hat{\beta}^{} \right)
= \, \left( \hat{c}^{\dagger} \,\, \hat{c}^{} \right) \,\,
\left(
\begin{array}{cc}
 U & V^{\star} \\
 V & U^{\star}
\end{array}
\right)
\end{equation}
or, equivalently:
\begin{equation}
 \left( \hat{\beta}^{\dagger} \,\, \hat{\beta}^{} \right)
= \, \left( \hat{\gamma}^{\dagger} \,\, \hat{\gamma}^{} \right) \,\,
\left(
\begin{array}{cc}
 \mathbb{U} & \mathbb{V}^{\star} \\
 \mathbb{V} & \mathbb{U}^{\star}
\end{array}
\right)
\end{equation}
where the matrices $\mathbb{U}$ and $\mathbb{V}$ are defined in \eqref{mathbbUV}.


Now, if we construct a family of operators $\hat{\beta}^{\prime}$ such that:
\begin{equation}
\hat{\beta}^{\prime}_l \, \exp\left( \hat{O} \right) = \exp\left( \hat{O} \right)  \hat{\beta}^{}_l
\end{equation}
then, by construction, the state $| \phi^{\prime} \rangle$ is the vacuum of this new family of operators since:
\begin{equation}
\hat{\beta}^{\prime}_l \, | \phi^{\prime} \rangle = \hat{\beta}^{\prime}_l \, \exp\left( \hat{O} \right) \, | \phi \rangle 
=  \exp\left( \hat{O} \right)  \hat{\beta}^{}_l \, | \phi \rangle = 0
\end{equation}
As shown in [\onlinecite{PhysRevB.95.045144}], 
the above considerations imply that
the operator $\exp\left( \hat{O} \right)$ transforms the operators:
\begin{equation}
\hat{\beta}^{}_l = 
\left( \,  \hat{c}^{\dagger} \, \,  \hat{c}^{} \right)
\,
\left(
\begin{array}{c}
 V^{\star}_{l}      \\
 U^{\star}_{l}
\end{array}
\right) = \, \left( \hat{\gamma}^{\dagger} \,\, \hat{\gamma}^{} \right) \,\,
\left(
\begin{array}{c}
 \mathbb{V}^{\star}_{l}      \\
 \mathbb{U}^{\star}_{l}
\end{array}
\right), \quad l=1, \dots, M
\end{equation}
into the new family of operators:
\begin{equation}
\label{hfbexpo:transf1}
\hat{\beta}^{\prime}_l = 
 \left( \,  \hat{c}^{\dagger} \, \,  \hat{c}^{} \right)
\, 
\,
\exp\left(
\begin{array}{cc}
 t  &  {\Delta}   \\
 \tilde{\Delta} & -t^{T}
\end{array}
\right)
\,
\left(
\begin{array}{c}
 V^{\star}_{l}      \\
 U^{\star}_{l}
\end{array}
\right)
\end{equation}
which shows that \eqref{exponential:map} maps $\mathcal{M}_{HFB}$ into itself.
In fact, it is simple to show 
that, if $\hat{O}$ is hermitian, the application of the exponential matrix conserves the conditions \eqref{hfbmanifold:HFB}, thus implying that \eqref{hfbexpo:transf1} is a fully legitimate Bogoliubov transformation.
The transformation \eqref{hfbexpo:transf1}, when ``rotated'' into the basis of the new vacuum gives: 
\begin{equation}
\label{hfbexpo:update}
\left(
\begin{array}{c}
 \mathbb{V}^{\prime \, \star}      \\
 \mathbb{U}^{\prime \, \star}
\end{array}
\right) = 
 \,\, e^{\mathcal{O}} \,\, \left(
\begin{array}{c}
 \mathbb{V}^{\star}      \\
 \mathbb{U}^{\star}
\end{array}
\right)
\end{equation}
where we introduced the notation:
\begin{equation}
\label{eq:def-expO}
e^{\mathcal{O}} \buildrel{def}\over{=} 
\left(
\begin{array}{cc}
 U^{\dagger}_v & V^{\dagger}_v \\
 V^T_v & U_v^T
\end{array}
\right)
\, 
\,
\exp\left(
\begin{array}{cc}
 t  &  {\Delta}   \\
 \tilde{\Delta} & -t^{T}
\end{array}
\right)
\,
\,\, \left(
\begin{array}{cc}
 U_v & V_v^{\star} \\
 V_v & U_v^{\star}
\end{array}
\right)
\end{equation}
This allows us to write 
$| \phi^{\prime} \rangle$ explicitly as:
\begin{equation}
\label{evolve_walker}
\begin{split}
& | \phi^{\prime} \rangle = \alpha \, 
e^{\hat{Z}^{\prime}} \, | {\omega}_v \rangle, 
\quad \hat{Z}^{\prime} = \sum_{\alpha < \beta} Z^{\prime}_{\alpha,\beta} \, \hat{\gamma}^{\dagger}_\alpha \hat{\gamma}^{\dagger}_\beta , \quad  Z^{\prime} = \left(  \mathbb{V}^{\prime} \mathbb{U}^{\prime \, -1} \right)^{\star}
\end{split}
\end{equation}
provided that the matrix $Z^{\prime}$ is well-defined.
Notice that, in \eqref{evolve_walker} we have a normalization constant $\alpha \in \mathbb{C}$, which can be determined by writing:
\begin{equation}
\exp\left( \hat{O} \right) \, | \phi \rangle = \exp\left( \hat{O} \right) \, e^{\hat{Z}} \, | {\omega}_v \rangle =  \alpha \, 
e^{\hat{Z}^{\prime}} \, | {\omega}_v \rangle
\end{equation}
The easiest way to calculate $\alpha$ explicitly is to compute the overlap of both sides of the above expression with $| {\omega}_v \rangle$, yielding,
\begin{equation}
\alpha = \langle \omega_v | \exp\left( \hat{O} \right) | \phi \rangle
\end{equation}
As detailed in Appendices A and B, we can compute this expression given $Z$ and the matrices $t$ and $\Delta$ that define $\hat{O}$. 
If we re-express Eq.~(\ref{eq:def-expO}) more compactly as
\begin{equation}
 e^{\mathcal{O}} \buildrel{def}\over{=}   
    \left(
    \begin{array}{cc}
    \mathbb{K} & \mathbb{M} \\
    \mathbb{L} & \mathbb{N}
    \end{array}
    \right)
\end{equation}
then we have:
\begin{equation}
\label{exponential:coefficient_alpha}
\begin{split}
   & \alpha = \sqrt{\det(\mathbb{N})} \, \left\langle
       \, \tilde{\psi} \big| \, \phi
    \right\rangle \\
    & | \tilde{\psi} \rangle \buildrel{def}\over{=}
    e^{\sum_{\alpha < \beta} \left( \mathbb{M} \mathbb{N}^{-1} \right) \,\hat{\gamma}^{\dagger}_{\alpha} \hat{\gamma}^{\dagger}_{\beta} } \, | \omega_v \rangle
\end{split}
\end{equation}
where the inner product can be computed with \eqref{overlap:pfaffian}.

\section{\label{sec:hfbmotion} Imaginary time dynamics in the manifold of Hartree-Fock-Bogoliubov Wave Functions}

Having established the essential mathematical and computational foundations of our approach, in this section
we focus on the final key ingredient of the method, which is the numerical implementation of imaginary time evolution
in the manifold of HFB wave functions.   

Consider a general mean-field hamiltonian of the form:
\begin{equation}
\label{hfbmotion:hatHmf}
\hat{H}_{\rm mf} =  \, \frac{1}{2} \, \left( \hat{c}^{\dagger} \,\, \hat{c}^{} \right) \,\, \mathcal{H}_{\rm mf} \,\, 
\left(
\begin{array}{c}
\hat{c}^{} \\
 \hat{c}^{\dagger}
\end{array}
\right)
\end{equation}
where $\mathcal{H}_{\rm mf}$ is a complex $2M \times 2M$ matrix 
\begin{equation}
\label{hfbmotion:Hmf}
\mathcal{H}_{mf} = \left(
\begin{array}{cc}
 t & \Delta \\
-\Delta^{\star} & -t^{T} 
\end{array}
\right)
\end{equation}
We denote the ground state of $\hat{H}_{\rm mf}$ in the Fock space of the physical system as $| \omega_{\rm mf} \rangle$ with corresponding energy $E_0$.

Our purpose in this section is to study the imaginary time dynamics underlying the projection formula:
\begin{equation}
\label{hfbmotion:projection}
| \omega_{\rm mf} \rangle \propto \lim_{\tau \to +\infty} \exp\left(- \tau \left(\hat{H}_{\rm mf} - E_0 \right) \right) \, |\phi_T \rangle
\end{equation}
in the special case when $|\phi_T \rangle$ (an HFB wave function referred to as the trial state) satisfies $\langle \omega_{\rm mf} | \phi_T \rangle \neq 0$.

The formalism described above is well-suited to treat systems with spin-polarized ground states, or, more generally, ground states with unpaired orbitals in $| \omega_{\rm mf} \rangle$. In these cases, it is convenient to rely on a ``new vacuum'' of the form:
\begin{equation}
| \omega_{v} \rangle \quad  \longleftrightarrow \quad \left(
\begin{array}{cc}
 U_v & V_v^{\star} \\
 V_v & U_v^{\star}
\end{array}
\right)
\end{equation}
Such an HFB wave function can be straightforwardly designed for a given physical problem and provides a natural parametrization to follow the path in $\mathcal{M}_{HFB}$, defined by \eqref{hfbmotion:projection}. 
Now, we choose our trial state $|\phi_T \rangle \in \mathcal{S}_v$:
\begin{equation}
| \phi_T \rangle \quad  \longleftrightarrow \quad \left(
\begin{array}{cc}
 U_T & V_T^{\star} \\
 V_T & U_T^{\star}
\end{array}
\right) \quad  \longleftrightarrow \left( \mathbb{U}_T, \mathbb{V}_T \right)
\end{equation}
and we express it as an unnormalized ``Thouless state'':
\begin{equation}
\begin{split}
& | \phi_T \rangle =  \, e^{\hat{Z}_{(T)}} \, | {\omega}_v \rangle \\
&
 \hat{Z}_{(T)} = \sum_{\alpha < \beta} Z_{T \, \alpha,\beta} \, \hat{\gamma}^{\dagger}_\alpha \hat{\gamma}^{\dagger}_\beta, \quad  Z_{T} = \left( \mathbb{V}_T \mathbb{U}_T^{-1}\right)^{\star}
\end{split}
\end{equation}
We note that it is always possible to choose $|\phi_T\rangle = |\omega_v\rangle$, in which case $Z_T = 0$, but we have the additional flexibility to allow the two wave functions to differ, which can enable very efficient calculations for correlated hamiltonians when a self-consistency scheme is implemented, as suggested in [\onlinecite{PhysRevB.94.235119}].

Put simply, we aim to carry out the imaginary time evolution in \eqref{hfbmotion:projection} by 
rotations in HFB space, which generate (non-orthogonal) excitations 
from the 
vacuum, $| {\omega}_v \rangle$, that contain information about the properties of the system, obtained from mean-field theory. The path \eqref{hfbmotion:projection} defines a family of wave functions in $\mathcal{M}_{HFB}$:
\begin{equation}
\label{hfbmotion:curve}
   \tau \to | \phi(\tau) \rangle = \exp\left(- \tau \left(\hat{H}_{mf} - E_0 \right) \right) \, |\phi_T \rangle
\end{equation}
which can be mapped onto a curve in the manifold of Bogoliubov matrices as follows:
\begin{equation}
\begin{split}
&
\left(
\begin{array}{c}
 \mathbb{V}^{ \star}(\tau + \delta \tau)      \\
 \mathbb{U}^{ \star}(\tau + \delta \tau)
\end{array}
\right) 
= \mathbb{B}_{\delta \tau}
\,
\left(
\begin{array}{c}
 \mathbb{V}^{\star}(\tau)      \\
 \mathbb{U}^{\star}(\tau)
\end{array}
\right), \\
&
\left(
\begin{array}{c}
 \mathbb{V}^{ \star}(\tau = 0)      \\
 \mathbb{U}^{ \star}(\tau = 0)
\end{array}
\right) 
= 
\left(
\begin{array}{c}
 \mathbb{V}_T^{ \star}     \\
 \mathbb{U}_T^{ \star}
\end{array}
\right) 
\end{split}
\end{equation}
where $\delta \tau$ is a time step. Using Eq.~\eqref{hfbexpo:update}
and Eq.~(\ref{eq:def-expO}), we denote:
\begin{equation}
     \mathbb{B}_{\delta \tau} \buildrel{def}\over{=} \left(
\begin{array}{cc}
 U^{\dagger}_v & V^{\dagger}_v \\
 V^T_v & U_v^T
\end{array}
\right)
\, 
\,
     \exp\left(
- \delta\tau \mathcal{H}_{mf}
\right)
\,
\,
\left(
\begin{array}{cc}
 U_v & V_v^{\star} \\
 V_v & U_v^{\star}
\end{array}
\right)
\end{equation}
The actual state:
\begin{equation}
| \phi(\tau) \rangle = \alpha(\tau) \,\, e^{\sum_{\alpha < \beta} Z_{\alpha,\beta}(\tau) \, \hat{\gamma}^{\dagger}_{\alpha} \hat{\gamma}^{\dagger}_{\beta}} \, | \omega_v \rangle
\end{equation}
can be constructed explicitly through the anti-symmetric matrix 
$Z(\tau) = \left( \mathbb{V}(\tau)  \mathbb{U}(\tau)^{-1}\right)^{\star}$
which allows us to compute all expectation values of physical observables using \eqref{rhok}-\eqref{sigmak}.  In addition, as proved in Appendices A and B, the coefficient $\alpha(\tau)$ can be computed using \eqref{exponential:coefficient_alpha}.
We stress that such a coefficient does not play any role in the imaginary time dynamics governed by a mean-field hamiltonian (as it cancels when we compute physical observables). On the other hand, $\alpha(\tau)$ will be crucial when we treat correlated hamiltonians in the following sections.

Along the imaginary-time path, different expectation values can be conveniently computed.
For example, with any quadratic operators 
\COMMENTED{
to obtain the pairing tensor:
\begin{equation}
 \tau_{\alpha \beta}(\tau) = \frac{\langle \phi(\tau) \, |  \hat{c}^{}_{\beta} \, \hat{c}^{}_{\alpha} \,  | \,  \phi(\tau) \rangle }{\langle \phi(\tau) \, | \, \phi(\tau) \rangle }, \quad \alpha, \beta = 1, \dots, M
\end{equation}
}
we only need to compute the density matrices in the algebra of the vacuum $| {\omega}_v \rangle$ using \eqref{rhok}-\eqref{sigmak} 
and then transform back into the original basis using \eqref{pairingtransformback}, 
as further illustrated below in the example.

A final issue concerns 
the numerical stabilization of these calculations. Consider the two matrices $\mathbb{U}(\tau)$ and $\mathbb{V}(\tau)$.
They form the transformation matrix between the Bogoliubov operators defining the HFB wave function, $|\phi(\tau)\rangle$, and those defining the new vacuum, $|\omega_v\rangle$. For this reason, they must satisfy the same conditions we have in \eqref{hfbmanifold:UVprop}.
In particular, the matrix:
$\mathbb{B}(\tau) = \mathbb{U}(\tau)^{T} \mathbb{V}(\tau)$
must be anti-symmetric. 

When computed numerically the matrices $\mathbb{U}(\tau)$ and $ \mathbb{V}(\tau)$ may contain round-off errors, which can result in the matrix $\mathbb{B}(\tau)$ no longer being anti-symmetric. We can ensure that $\mathbb{B}(\tau)$ is anti-symmetric by imposing the condition, $\mathbb{B}_{\alpha,\beta}(\tau) = - \mathbb{B}_{\beta,\alpha}(\tau)$ if $\alpha \geq \beta$ and then redefining:
$\mathbb{V}(\tau) \rightarrow   \mathbb{U}(\tau)^{-1, \, T} \mathbb{B}(\tau) $
as $\mathbb{U}(\tau)$ is invertible by construction within $\mathcal{S}_v$.
At this point, we can perform a Gram-Schmidt procedure (QR decomposition):
\begin{equation}
\left(
\begin{array}{c}
 \mathbb{U}(\tau)   \\
   \mathbb{V}(\tau)
\end{array}
\right)
= 
\left(
\begin{array}{c}
 \tilde{\mathbb{U}}(\tau)  \\
  \tilde{\mathbb{V}}(\tau)
\end{array}
\right)
\mathbb{R}(\tau)
\end{equation}
where $\left(
\begin{array}{c}
 \tilde{\mathbb{U}}(\tau)   \\
  \tilde{\mathbb{V}}(\tau)
\end{array}
\right)$ is $2M \times M$ with orthonormal columns, while $\mathbb{R}(\tau)$ is an $M \times M$ upper triangular matrix. Notice that:
\begin{equation}
Z(\tau) = \left( \mathbb{V}(\tau) \, \mathbb{U}^{-1}(\tau) \right)^{\star} =  \left( \tilde{\mathbb{V}}(\tau) \, \tilde{\mathbb{U}}^{-1}(\tau) \right)^{\star} 
\end{equation}
so that $Z(\tau)$ is not affected by the stabilization. Such a stabilization procedure mirrors what is routinely done with Slater determinants within AFQMC [\onlinecite{https://doi.org/10.1002/wcms.1364}].

\subsection{\label{sec:hfbillustration} Illustrative examples for mean-field HFB hamiltonians}

\begin{figure}
    \centering
    \includegraphics[width=1\linewidth]{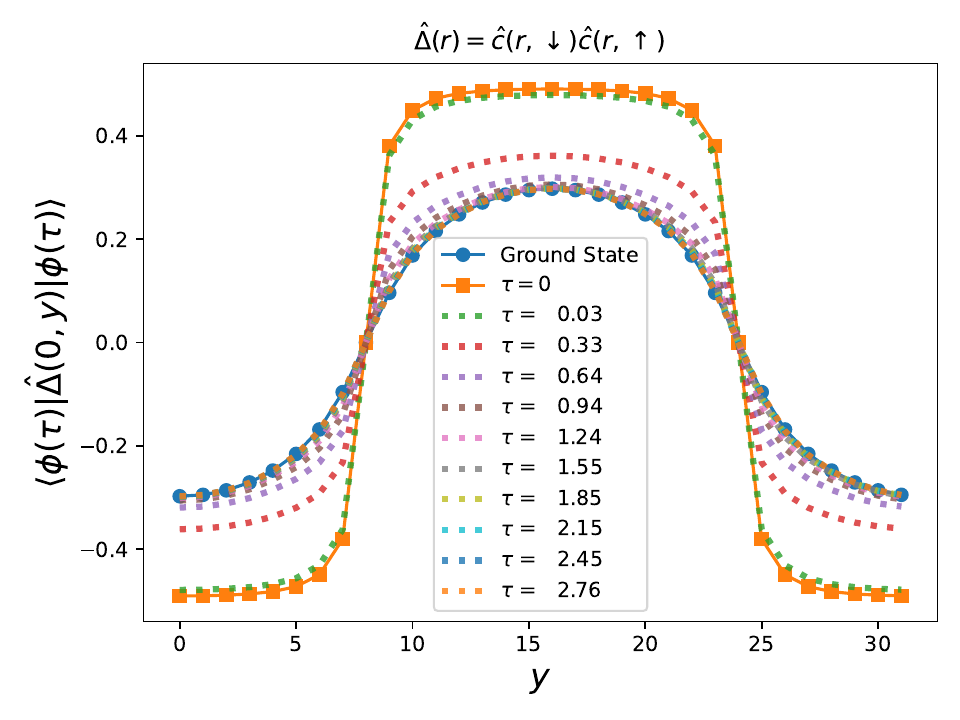}
    \caption{Imaginary time dependence of the on-site pairing order parameter $\tau_{r,\downarrow; r,\uparrow}(\tau)$ 
    defined in \eqref{hfbmotion:tauruprdn} on a $4 \times 32$ lattice. The blue circles show the expectation value over $|\omega_{\rm mf}\rangle$ in \eqref{hfbmotion:projection} chosen to be the ground state of $\hat{H}_{\rm mf}$ with $t=1$, $\mu=0$, $\lambda=0$, $h_{\sigma}(r) = h_0 \, (-1)^{\sigma}$, $h_0 = 0.1$ and $\eta(r=(x,y)) = \eta_0 \cos( \pi y /16 )$, with $\eta_0 = 1$ (\eqref{meanfield:matrixt} and \eqref{meanfield:matrixD}).
    The trial state $|\phi_T \rangle$ is obtained by changing the value of $\eta_0$ to $10$, and the corresponding pairing order parameter is shown as orange squares. The dotted lines are the results for finite imaginary time $\tau$.}
    \label{fig:illustration_hfb_homemade}
\end{figure}

As a first 
illustration of the formalism described above, we focus on two-dimentional lattice models for spin $1/2$ fermions, and use the basis $|r,\sigma\rangle$, where the label $r$ runs over the $L$ sites of a finite square lattice while $\sigma = \uparrow, \downarrow$. We choose a fairly general form for the hermitian matrix in \eqref{hfbmotion:Hmf}:
\begin{equation}
\label{meanfield:matrixt}
\begin{split}
& t_{r,\sigma ; r^{\prime},\sigma^{\prime}}
=  \delta_{\sigma, \sigma^{\prime}} ( t^{hop}_{r,r^{\prime}} - \mu \delta_{r,r^{\prime}}) \\
& 
- \lambda \delta_{\sigma, -\sigma^{\prime}} \, t^{soc}_{r,r^{\prime}}
- \delta_{\sigma, \sigma^{\prime}} \delta_{r, r^{\prime}} h_{\sigma}(r)
\end{split}
\end{equation}
where $t^{hop}$ is the usual Hubbard model nearest-neighbor hopping matrix, 
with amplitude $-t$ if $r$ and $r^{\prime}$ 
are nearest neighbors and zero otherwise, and
the Rashba spin-orbit coupling term, controlled by an overall strength
$\lambda$, is 
defined by the same matrix $t^{soc}_{r,r^{\prime}}$ given in [\onlinecite{PhysRevLett.119.265301}].
In addition, $\mu$ is a chemical potential, 
while the function $h_{\sigma}(r)$ models external fields coupled to the density and/or the spin-density, for example a magnetic field favoring a spin-polarization.
We also choose 
the simplest form of the anti-symmetric matrix $\Delta$, specifically a singlet on-site pairing term:
\begin{equation}
\label{meanfield:matrixD}
    \Delta_{r,\sigma; r^{\prime},\sigma^{\prime}}
    = \delta_{r,r^{\prime}} \eta(r) \frac{1}{2} \left(\delta_{\sigma, \uparrow}\delta_{\sigma^{\prime}, \downarrow}
    - \delta_{\sigma, \downarrow}\delta_{\sigma^{\prime}, \uparrow}\right)\,.
\end{equation}
In Fig.~\ref{fig:illustration_hfb_homemade} we consider a $4 \times 32$ lattice with periodic boundary conditions. We choose $|\omega_{\rm mf}\rangle$ in \eqref{hfbmotion:projection} (referred to as ``Ground State'' in the figure) as the ground state of \eqref{hfbmotion:hatHmf} with $t=1$, $\mu=0$, $\lambda=0$ and a magnetic field imposing a spin polarization $h_{\sigma}(r) = h_0 \, (-1)^{\sigma}$, with $h_0 = 0.1$. In addition, we use a modulated pairing field of the form $\eta(r=(x,y)) = \eta_0 \cos( \pi y /16 )$, with $\eta_0 = 1$, in order to model a Fulde-Ferrell-Larkin-Ovchinnikov phase. The hamiltonian $\hat{H}_{\rm mf}$ with this choice of parameters is also used to drive the imaginary time dynamics in \eqref{hfbmotion:projection}. 

In our calculations, 
we choose the new vacuum $|\omega_v \rangle$ as the ground state of an hamiltonian of the same form as $\hat{H}_{\rm mf}$, but with a different value of the amplitude of the pairing term, in this case $\eta_0 = 10$. In this first example, we choose the trial state to be the vacuum itself $|\phi_T\rangle = |\omega_v \rangle$.

In order to test our methodology we implement the imaginary time dynamics in the HFB manifold as defined in \eqref{hfbmotion:curve} and compute the on-site terms of the pairing tensor:
\COMMENTED{
\begin{equation}
\label{hfbmotion:tauruprdn}
 \tau_{r,\uparrow ; r,\downarrow}(\tau) = \frac{\langle \phi(\tau) \, |    \hat{\Delta}(r)| \,  \phi(\tau) \rangle }{\langle \phi(\tau) \, | \, \phi(\tau) \rangle }, \quad
\hat{\Delta}(r) \buildrel{def}\over{=} \hat{c}^{}_{r,\downarrow} \, \hat{c}^{}_{r,\uparrow} \,
\end{equation}
}
\begin{equation}
\label{hfbmotion:tauruprdn}
 \tau_{r,\uparrow ; r,\downarrow}(\tau) = \frac{\langle \phi(\tau) \, |    
 \hat{c}^{}_{r,\downarrow} \, \hat{c}^{}_{r,\uparrow}
 | \,  \phi(\tau) \rangle }{\langle \phi(\tau) \, | \, \phi(\tau) \rangle }\,, 
\end{equation}
as functions of the imaginary time $\tau$. The value at $\tau=0$ (orange squares in the figure) is the expectation value over $|\phi_T\rangle$, while the expectation value over $|\omega_{mf}\rangle$ is plotted as blue circles. The dotted lines show the results for \eqref{hfbmotion:tauruprdn} at finite $\tau$. The figure convincingly shows the convergence to the ground state expectation value with no evidence of significant round-off errors.

\begin{figure}
    \centering
    \includegraphics[width=1\linewidth]{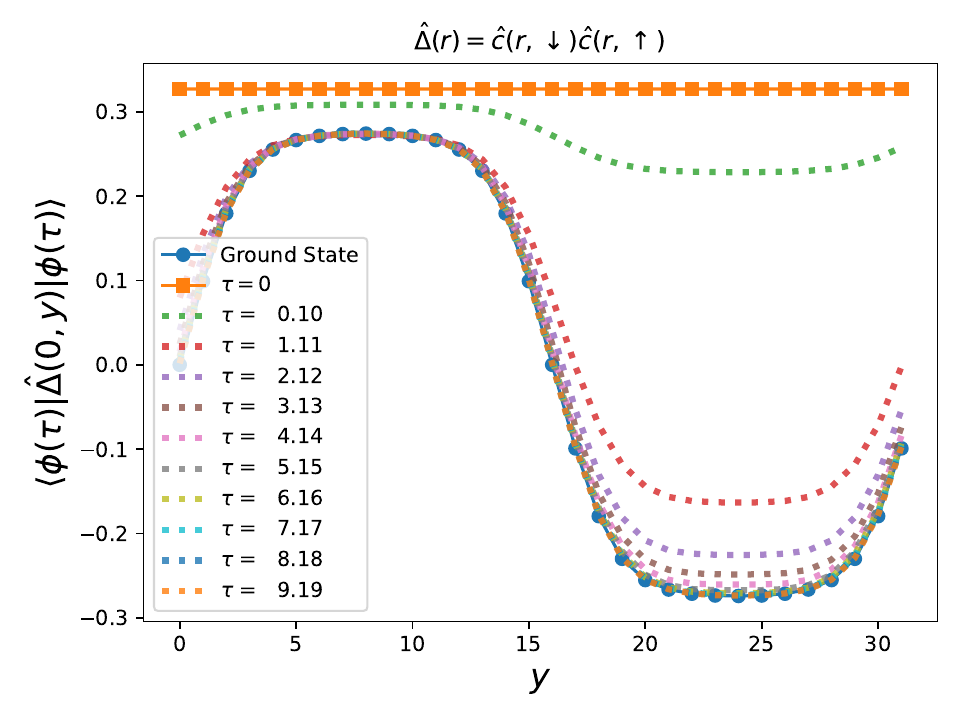}
    \caption{Imaginary time dependence of the on-site pairing order parameter $\tau_{r,\downarrow; r,\uparrow}(\tau)$ 
    defined in \eqref{hfbmotion:tauruprdn} on a $4 \times 32$ lattice. The blue circles show the expectation value over $|\omega_{\rm mf}\rangle$ in \eqref{hfbmotion:projection} chosen to be the fully self-consistent HFB solution with $U=-2.75$ in \eqref{meanfieldbreakup}, while $\mu$, $h_\sigma(r)$ and $\eta(r)$ in \eqref{meanfield:matrixt} and \eqref{meanfield:matrixD} are determined self-consistently using \eqref{meanfieldbreakup} by choosing average particle density $\langle n \rangle =0.95$ and average spin polarization $p = \frac{1}{2}(\langle n_{\uparrow} \rangle - \langle n_{\downarrow} \rangle ) = 0.03125$. The trial state $|\phi_T \rangle$ is obtained by changing only the value of $U$ to $-4$, and the corresponding pairing order parameter is shown as orange squares. The dotted lines are the results for finite imaginary time $\tau$.}
    \label{fig:ch1_projection_hfb}
\end{figure}

As a second illustrative example, we consider fully self-consistent Hartree-Fock-Bogoliubov calculations, 
applying a mean-field break-up of the interaction term of the form:
\begin{equation}
\label{meanfieldbreakup}
\begin{split}
& \hat{V} = U \sum_{r} \hat{c}^{\dagger}_{r,\uparrow} \hat{c}^{}_{r,\uparrow}
\hat{c}^{\dagger}_{r,\downarrow} \hat{c}^{}_{r,\downarrow} 
\Rightarrow \\
&  U \sum_{r} \langle \hat{c}^{\dagger}_{r,\uparrow} \hat{c}^{}_{r,\uparrow} \rangle 
\hat{c}^{\dagger}_{r,\downarrow} \hat{c}^{}_{r,\downarrow} +
 U \sum_{r}  \hat{c}^{\dagger}_{r,\uparrow} \hat{c}^{}_{r,\uparrow}  
\langle \hat{c}^{\dagger}_{r,\downarrow} \hat{c}^{}_{r,\downarrow} \rangle \\
& +  U \sum_{r} \langle \hat{c}^{\dagger}_{r,\uparrow} \hat{c}^{\dagger}_{r,\downarrow} \rangle
 \hat{c}^{}_{r,\downarrow} \hat{c}^{}_{r,\uparrow} +  U \sum_{r}  \hat{c}^{\dagger}_{r,\uparrow} \hat{c}^{\dagger}_{r,\downarrow}  
\langle \hat{c}^{}_{r,\downarrow} \hat{c}^{}_{r,\uparrow} \rangle\,. 
\end{split}
\end{equation}
We consider again a $4 \times 32$ lattice with cylindrical boundary conditions, we still choose $t=1$, but now consider a finite spin-orbit coupling 
strength $\lambda = 0.05$. We choose $|\omega_{\rm mf} \rangle$ (and $\hat{H}_{\rm mf}$) by letting $U=-2.75$ in \eqref{meanfieldbreakup}, while $\mu$, $h_\sigma(r)$ and $\eta(r)$ in \eqref{meanfield:matrixt} and \eqref{meanfield:matrixD} are determined self-consistently using \eqref{meanfieldbreakup} by choosing average particle density $\langle n \rangle =0.95$ and average spin polarization $p = \frac{1}{2}(\langle n_{\uparrow} \rangle - \langle n_{\downarrow} \rangle ) = 0.03125$.
For the purpose of illustration, we choose the new vacuum $|\omega_v \rangle$ by repeating the procedure only changing $U$ to $U=-2$, while $|\phi_T \rangle$ corresponds to $U=-4$. This is a challenging test since, as is evident from Fig.~\ref{fig:ch1_projection_hfb}, which shows that the trial state has a constant pairing order parameter, while the ground state has a modulated one. 
As seen, the projection leads to 
convergence to the ground state expectation value with no sign of significant round-off errors, even for large values of the imaginary time.

\section{\label{sec:qmcmotion} Correlated Hamiltonians}
The results in the previous section verify the feasibility and numerical efficiency of our framework to carry out imaginary time evolution in the manifold of HFB wave functions, even in the most general situation involving unpaired fermions, i.e. situations in which the structure of the pairing cannot be constrained as in [\onlinecite{PhysRevA.100.023621}]. With respect to the usual implementations relying on Slater determinants with $N$ particles, where the central object is an $M \times N$ matrix of spin orbitals,  in the HFB manifold we need to manipulate two $M \times M$ matrices, $\mathbb{U}$ and $\mathbb{V}$ and the corresponding matrix $Z(\tau)$. All of the operations used in the previous section to compute the pairing tensor involve matrix multiplications and inversions, leading to a computational complexity $\mathcal{O}(M^3)$, equivalent (although likely with a larger prefactor) to finite temperature determinantal calculations. 
The most computationally expensive operation, which is relevant only for correlated hamiltonians, is the calculation of the pfaffian of a $2M \times 2M$ anti-symmetric matrix in \eqref{overlap:pfaffian}. 
Efficient techniques have been developed to compute the pfaffian 
[\onlinecite{10.1145/2331130.2331138, PhysRevB.77.115112, doi:10.1143/JPSJ.77.114701}] with a cubic scaling in $M$, but it remains an important question whether the advantage of working with HFB wave functions that contain information about pairing physics is outweighed by the additional computational cost of computing pfaffians as opposed to determinants. In addition, we note that, since $pf(A)^2 = \det(A)$ for any anti-symmetric matrix, only the sign of the overlap needs to be determined in \eqref{overlap:pfaffian}, while the absolute value can be determined by simply computing a determinant.

We have now all the ingredients to explore the possibility of treating general quartic hamiltonians, as in 
electronic problems with 
Coulomb interactions. 
For simplicity, we use a Hubbard interaction in our illustration below:
\begin{equation}
\label{correlatedham}
\hat{H} = \hat{H}_{\rm mf} + \hat{V}, \quad  \hat{V} = U \sum_{r} \hat{c}^{\dagger}_{r,\uparrow} \hat{c}^{}_{r,\uparrow}
\hat{c}^{\dagger}_{r,\downarrow} \hat{c}^{}_{r,\downarrow}\,.
\end{equation}
We consider a 
Hubbard interaction with $U < 0$ to highlight the 
utility of HFB, while $\hat{H}_{\rm mf}$ has the structure described in the previous section.
We would like to use the same projection formula \eqref{hfbmotion:projection} with the correlated hamiltonian, to find the ground state:
\begin{equation}
\label{qmcmotion:projection}
| \Psi_0 \rangle \propto \lim_{\tau \to +\infty} \exp\left(- \tau \left(\hat{H} - E_0 \right) \right) \, |\phi_T \rangle
\end{equation}
where, as before, we choose $|\phi_T\rangle \in \mathcal{M}_{HFB}$.
The well-known charge decomposition Hubbard Stratonovich transformation:
\begin{equation}
\begin{split}
& e^{-\delta \tau U \, \hat{c}^{\dagger}_{r,\uparrow} \hat{c}^{}_{r,\uparrow}
\hat{c}^{\dagger}_{r,\downarrow} \hat{c}^{}_{r,\downarrow}}
= e^{- \delta \tau U \left(\hat{c}^{\dagger}_{r,\uparrow}\hat{c}^{}_{r,\uparrow} + \hat{c}^{\dagger}_{r,\downarrow}\hat{c}^{}_{r,\downarrow} - 1\right) / 2} \\
& 
\sum_{x = \pm 1} \frac{1}{2} e^{\gamma x \left(\hat{c}^{\dagger}_{r,\uparrow}\hat{c}^{}_{r,\uparrow} + \hat{c}^{\dagger}_{r,\downarrow}\hat{c}^{}_{r,\downarrow} - 1\right)}
\end{split}
\end{equation}
can be expressed in the following form, more suitable for our formalism:
\begin{equation}
e^{-\delta \tau U \, \hat{c}^{\dagger}_{r,\uparrow} \hat{c}^{}_{r,\uparrow}
\hat{c}^{\dagger}_{r,\downarrow} \hat{c}^{}_{r,\downarrow}}
= \sum_{x = \pm 1} \frac{1}{2 }  e^{ \frac{1}{2} \left( \hat{\bf{c}}^{\dagger} \,\, \hat{\bf{c}}^{}   \right) \,
\mathcal{O}_V(r,x) \,
\left( 
\begin{array}{c}
\hat{\bf{c}}^{} \\
\hat{\bf{c}}^{\dagger}
\end{array}
\right)  }
\end{equation}
In the above expression we have introduced the $2M \times 2M$ diagonal matrix:
\begin{equation}
\mathcal{O}_V(r,x) \ = 
\left( 
\begin{array}{cccc}
\mathcal{O}^{\uparrow\uparrow}_V(r,x)  & 0 & 0 & 0 \\
0 & \mathcal{O}^{\downarrow\downarrow}_V(r,x) & 0 & 0 \\
0 & 0 & -  \mathcal{O}^{\uparrow\uparrow}_V(r,x) & 0 \\
0 & 0 & 0 & - \mathcal{O}^{\downarrow\downarrow}_V(r,x)
\end{array}
\right)
\end{equation}
with:
\begin{equation}
\mathcal{O}^{\uparrow\uparrow}_V(r,x)_{r_1, r_2} = \mathcal{O}^{\downarrow\downarrow}_V(r,x)_{r_1, r_2} = \delta_{r_1, r_2} \delta_{r_1, r} 
\left( - \left( \frac{\delta \tau U}{2} - \gamma x\right) \right)
\end{equation}

Now, if we introduce a Trotter decomposition, we can express:
\begin{equation}
\label{afqmchfb}
\begin{split}
& e^{-\delta \tau \hat{H}} = e^{ \frac{1}{2} \left( \hat{\bf{c}}^{\dagger} \,\, \hat{\bf{c}}^{}   \right) \,
\left( - \delta\tau \mathcal{H}_{mf}  \right) \,
\left( 
\begin{array}{c}
\hat{\bf{c}}^{} \\
\hat{\bf{c}}^{\dagger}
\end{array}
\right)  } \\
& \prod_{r} \sum_{x(r)} e^{ \frac{1}{2} \left( \hat{\bf{c}}^{\dagger} \,\, \hat{\bf{c}}^{}   \right) \,
\mathcal{O}_V(r,x(r)) \,
\left( 
\begin{array}{c}
\hat{\bf{c}}^{} \\
\hat{\bf{c}}^{\dagger}
\end{array}
\right)  } \, \, + \mathcal{O}(\delta\tau)\,.
\end{split}
\end{equation}
All the operators 
are now in forms which can act within $\mathcal{M}_{HFB}$ following the same procedure discussed in the previous section.
We stress that the above expression is perfectly analogous to the imaginary time evolution in the manifold of $N$-particle Slater determinants that underlies the well known auxiliary-field quantum Monte Carlo method [\onlinecite{PhysRevB.55.7464},\onlinecite{PhysRevLett.90.136401}]. 
Formula \eqref{afqmchfb} ``promotes'' AFQMC to the manifold $\mathcal{M}_{HFB}$. In the previous section we developed the formalism needed to implement the new random walk within $\mathcal{M}_{HFB}$.
In the language of open-ended random walks in AFQMC, we consider a collection of walkers, which are now elements of $\mathcal{M}_{HFB}$. Each walker is parametrized by two $M \times M$ matrices, $\mathbb{U}$ and $\mathbb{V}$ (and the corresponding matrix $Z$), and by a coefficient $\alpha$ as in \eqref{exponential:coefficient_alpha}, which is now crucial as it depends on the auxiliary-field configuration.
All of the calculations possible within determinantal AFQMC can be performed with this approach, using the density matrices \eqref{rhok}-\eqref{sigmak} for one-body operators and leveraging Wick's theorem (which holds within $\mathcal{M}_{HFB}$) for two-body operators. 
The major advantage with this formalism is that we can study hamiltonians that do not conserve particle number, and we can compute pairing order parameters, which are crucial to characterize superfluids and superconductors. In fact, the ability to introduce into the hamiltonian a term that does not conserve the number of particles allows us to study the response to a pairing field, which provides a way to directly compute pairing order parameters. When the order parameters are small, this provides a significant advantage over the usual approach of computing two-body correlation functions, which yield the square of the (often tiny) order parameter. 

We also comment that, as in AFQMC, we need to implement a constrained-path approximation by allowing only walkers non-orthogonal to $|\omega_v\rangle$ 
[\onlinecite{PhysRevB.55.7464}], and it seems natural to expect that the flexibility in the choice of the new vacuum can be a huge asset to minimize any bias.

\subsection{Comparison with exact diagonalization}
As a test of robustness of the methodology, we consider a $2 \times 2$ lattice, which can be treated using exact diagonalization of \eqref{correlatedham} in the $256$-dimensional Fock space.
We focus on three values of the interaction strength $U=-2.8 t$, $U=-6.0 t$ and $U=-10.0 t$. We choose $\hat{H}_{mf}$ by letting $t=1$, $\mu = U/2$, $h_{\sigma}(r) = h_0 (-1)^{\sigma}$ with $h_0 = 1.5$, and $\eta(r)=\eta_0 = -0.5$.

\begin{figure}
    \centering
    \includegraphics[width=1\linewidth]{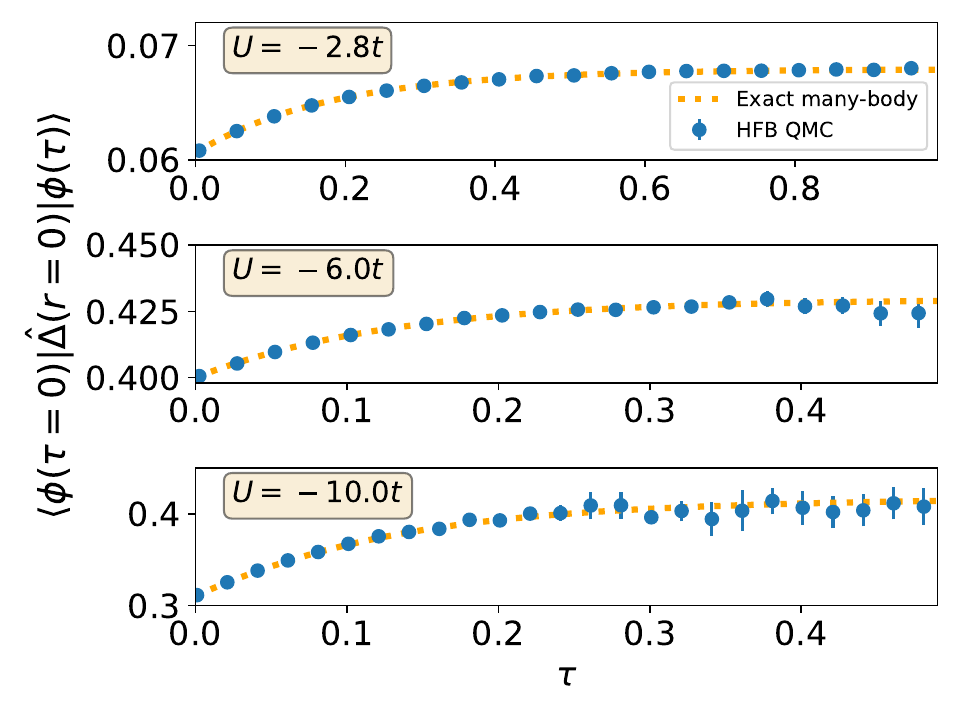}
    \caption{Imaginary time dependence of the mixed estimator of the on-site pairing order parameter $\tau_{r=0,\downarrow; r=0,\uparrow}(\tau) = \langle \phi(\tau=0) | \hat{c}^{}_{0,\downarrow} \hat{c}^{}_{0,\uparrow} | \phi(\tau) \rangle$ for the hamiltonian \eqref{correlatedham} with $B(r) = B = 1.5 \, t$, $\eta(r) = \eta = -0.5$, and three different values of $U$: $U = -2.8 \, t$ (upper panel), $U = -6.0 \, t$ (middle panel), 
    $U = -10.0 \, t$ (lower panel). In all cases we set $\mu=U/2$. Blue circles are results from our method (error bars, where not evident, are smaller than the symbol size), dotted line is obtained with exact diagonalization.
    }
\label{fig:2b2_projection_qmc}
\end{figure}

The value $U=-2.8 t$, with our choice of magnetic field and pairing term, corresponds to the largest value of interaction strength such that the overlap between the correlated ground state and the non-interacting ground state does not vanish. In such a case, we can use as a new vacuum and trial state the ground state $|\omega_{mf} \rangle$ of \eqref{correlatedham} with $U=0$, and implement an open-ended random walk sampling the imaginary time dynamics governed by \eqref{afqmchfb} as described above.
In the upper panel of Fig.~\ref{fig:2b2_projection_qmc} we show the result for the mixed estimator of the on-site pairing order parameter:
\begin{equation}
    \frac{\langle \phi(\tau=0) | \hat{c}^{}_{r,\downarrow} \hat{c}^{}_{r,\uparrow} \, | \, \phi(\tau) \rangle}{\langle \phi(\tau=0) |  \phi(\tau) \rangle}
\end{equation}
The dotted lines are obtained using exact diagonalization, while the blue circles with error-bars are obtained with AFQMC in the manifold of HFB wave functions. The pure estimators can be obtained as well by using techniques analogous to those used in standard 
AFQMC, for example back propagation [\onlinecite{10.1063/1.5029508}] or dynamical evolution of operators, as discussed in [\onlinecite{PhysRevA.100.023621}]. We obtain excellent agreement with exact diagonalization, which 
indicates the promise 
of our approach.

For larger values of the interaction strength, $|\omega_{mf} \rangle$ is orthogonal to the correlated ground state, and so it not possible to use it as a vacuum or trial state. 
For the problem we consider here, this orthogonality is due to the fact that the non-interacting state is spin-polarized, whereas the correlated state, despite the presence of a magnetic field, is spin-balanced, due to the attractive interaction energy. In such cases, we can leverage the flexibility to choose a new vacuum state, which we construct by diagonalizing $\hat{H}_{mf}$ with $B_0=0$ and $\eta_0 = -1.5$ for $U=-6t$ and  $B_0=0$ and $\eta_0 = -0.5$ for $U=-10t$. We comment that the value of $\eta_0$ can be used to fine tune the starting point of the imaginary time dynamics, as can be seen from the middle and bottom panels of Fig.~\ref{fig:2b2_projection_qmc} at $\tau = 0$.   
The comparison with exact diagonalization again shows excellent agreement, even at $U=-6.0 t$ (middle panel) and $U=-10.0 t$ (lower panel), where there are larger fluctuations due to dominant contribution of the interaction term. We also comment that self-consistency loops can be used to systematically optimize the choice of the vacuum state by leveraging the technique proposed in [\onlinecite{PhysRevB.94.235119}].


\section{\label{sec:conclusion} Conclusions and Perspectives}
We propose a 
general quantum Monte Carlo methodology designed to treat a broad range of physical systems, especially ones
in which paring is present,
including spin-imblalanced fermions subject to spin-orbit coupling. The framework we develop carries out a random walk in the manifold of Hartree-Fock-Bogoliubov wave functions, which extends the state-of-the-art 
AFQMC method that is typically realized as a random walk in the manifold of Slater determinants. By leveraging the flexibility to choose a vacuum state already containing important information about the pairing mechanisms of the system, we design an approach that can treat correlations beyond the level of mean field theory. As we demonstrate, this can be achieved by stochastically 
exciting quasi-particles
via rotations in HFB space, in such a way that statistical averages yield ground state correlations.
We showcase the possibility to implement imaginary time dynamics of HFB states in a numerically efficient and stable way.
Having established the feasibility of our method, we plan to test the approach for larger systems and explore different potential applications, including a study of the two-dimensional Hubbard model with spin-orbit coupling and Zeeman fields, a model that can be realized in cold atom systems and is expected to support exotic phases with topological features.
Another potential direction for future work is to extend the formalism to treat systems at finite temperature, which will enable direct comparison with cold atom experiments.

\begin{acknowledgments}
We acknowledge valuable discussions with Hao Shi. E.V. is supported by the National Science Foundation (Award number PHY-2207048).
The Flatiron Institute is a division of the Simons Foundation.
\end{acknowledgments}

\appendix

\section{Vacuum expectation value of exponentials of hermitian quadratic operators}
In this Appendix, we consider an hermitian operator of the form:
\begin{equation}
  \hat{O} =    \frac{1}{2}  \left( \hat{\bf{\gamma}}^{\dagger} \,\, \hat{\bf{\gamma}}^{}   \right) \,
\left( \mathcal{O}  \right) \,
\left( 
\begin{array}{c}
\hat{\bf{\gamma}}^{} \\
\hat{\bf{\gamma}}^{\dagger}
\end{array}
\right)
\end{equation}
defined by a $2M \times 2M$ matrix with the usual redundant form:
\begin{equation}
 \mathcal{O} \buildrel{def}\over{=} 
 \left(
    \begin{array}{cc}
    \mathcal{T} & \mathcal{A} \\
    -\mathcal{A}^{\star} & - \mathcal{T}^{T}
    \end{array}
    \right)
\end{equation}
where $\mathcal{T}$ is an hermitian $M \times M$ matrix and
$\mathcal{A}$ is an anti-symmetric $M \times M$ matrix.
As we did in the main text, we focus on a vacuum $| \omega_v \rangle$ with its algebra $\hat{\gamma^{}}, \hat{\gamma}^{\dagger}$.
It will be useful to introduce the notation:
\begin{equation}
    \exp\left(\mathcal{O}\right)
    \buildrel{def}\over{=} 
    \left(
    \begin{array}{cc}
    \mathbb{K} & \mathbb{M} \\
    \mathbb{L} & \mathbb{N}
    \end{array}
    \right)
\end{equation}
where all the blocks are $M \times M$ complex matrices.
We are going to show that:
\begin{equation}
\label{AppendixA:ExpectationExpO}
   \left\langle  \omega_v \, | \, \exp(\hat{O})  \, | \, \omega_v \right\rangle
   = \sqrt{\det(\mathbb{N})} 
\end{equation}

In order to prove this important theorem, following Hara and Iwasaki \cite{HARA197961}, we go back to the transformation formula:
\begin{equation}
\label{Appendix:conjugatetrans}
\hat{\beta}^{\prime}_l \,  = \exp\left( \hat{O} \right)  \hat{\beta}^{}_l \, \exp\left(- \hat{O} \right)
\end{equation}
where we consider a set of quasi-particle operators:
\begin{equation}
\hat{\beta}^{}_l = \, \left( \hat{\gamma}^{\dagger} \,\, \hat{\gamma}^{} \right) \,\,
\left(
\begin{array}{c}
 \mathbb{V}^{\star}_{l}      \\
 \mathbb{U}^{\star}_{l}
\end{array}
\right), \quad l=1, \dots, M
\end{equation}
Now, using the well known expansion for the conjugation:
\begin{equation}
    \exp\left( \hat{O} \right)  \hat{\gamma}^{} \, \exp\left(- \hat{O} \right) = \hat{\gamma}^{} + \left[ \hat{O} \, , \, \hat{\gamma}^{} \right] + \frac{1}{2!} \left[ \hat{O} \, , \, \left[ \hat{O} \, , \, \hat{\gamma}^{} \right] \right] + \dots 
\end{equation}
it is straightforward to get to the formula:
\begin{equation}
    \exp\left( \hat{O} \right)  \, \left( \hat{\gamma}^{\dagger} \,\, \hat{\gamma}^{} \right) \, \exp\left(- \hat{O} \right)
    \,\, \left(
\begin{array}{c}
 \mathbb{V}^{\star}_{l}      \\
 \mathbb{U}^{\star}_{l}
\end{array}
\right)
= \left( \hat{\gamma}^{\dagger} \,\, \hat{\gamma}^{} \right)
\, \exp\left(\mathcal{O}\right) \,  \,\, \left(
\begin{array}{c}
 \mathbb{V}^{\star}_{l}      \\
 \mathbb{U}^{\star}_{l}
\end{array}
\right)
\end{equation}

Now, let us consider the special case:
\begin{equation}
    \hat{\beta}_l = \hat{\gamma}_l, \quad
 \left( \begin{array}{c}
 \mathbb{V}^{\star}_{l}      \\
 \mathbb{U}^{\star}_{l}
\end{array} 
 \right)
=
 \left(
\begin{array}{c}
 0      \\
 \dots \\
 0 \\
 0 \\
 \dots \\
 0 \\
 1 \\
 0 \\
 \dots \\
 0 \\
\end{array}
\right)
\end{equation}
We find:
\begin{equation}
    \exp\left( \hat{O} \right)  \, \, \hat{\gamma}^{}_l  \, \exp\left(- \hat{O} \right)
= \sum_{j=1}^{M} \hat{\gamma}^{\dagger}_{j} \, \mathbb{M}_{jl}
+ \sum_{j=1}^{M} \hat{\gamma}^{}_{j} \, \mathbb{N}_{jl}
\end{equation}
An analogous calculation leads to:
\begin{equation}
    \exp\left( \hat{O} \right)  \, \, \hat{\gamma}^{\dagger}_l  \, \exp\left(- \hat{O} \right)
= \sum_{j=1}^{M} \hat{\gamma}^{\dagger}_{j} \, \mathbb{K}_{jl}
+ \sum_{j=1}^{M} \hat{\gamma}^{}_{j} \, \mathbb{L}_{jl}
\end{equation}
Introducing a parameter $\theta \in \mathbb{R}$,
and denoting:
\begin{equation}
\label{Appendix:Wtheta}
    \mathcal{W}(\theta) \buildrel{def}\over{=}
    \exp\left( \theta \mathcal{O} \right)
    \buildrel{def}\over{=} \left(
    \begin{array}{cc}
    \mathbb{K}(\theta) & \mathbb{M}(\theta) \\
    \mathbb{L}(\theta) & \mathbb{N}(\theta)
    \end{array}
    \right)
\end{equation}
we can write:
\begin{equation}
\label{Appendix:conjugationKLMN}
    \begin{split}
        & \exp\left( \theta\hat{O} \right)  \, \, \hat{\gamma}^{}_l  \, \exp\left(-\theta \hat{O} \right)
= \sum_{j=1}^{M} \hat{\gamma}^{\dagger}_{j} \, \mathbb{M}_{jl}(\theta)
+ \sum_{j=1}^{M} \hat{\gamma}^{}_{j} \, \mathbb{N}_{jl}(\theta) \\
& \exp\left( \theta\hat{O} \right)  \, \, \hat{\gamma}^{\dagger}_l  \, \exp\left(- \theta\hat{O} \right)
= \sum_{j=1}^{M} \hat{\gamma}^{\dagger}_{j} \, \mathbb{K}_{jl}(\theta)
+ \sum_{j=1}^{M} \hat{\gamma}^{}_{j} \, \mathbb{L}_{jl}(\theta)
    \end{split}
\end{equation}
Now, let us introduce two operators $\hat{O}_1$ and $\hat{O}_2$, and denote $\langle \dots \rangle \buildrel{def}\over{=} \langle \omega_v | \dots | \omega_v \rangle$; the relations \eqref{Appendix:conjugationKLMN} imply immediately:
\begin{equation}
\label{Applendix:O1O2first}
    \left\langle   \, \hat{O}_1 e^{ \theta\hat{O} }  \, \, \hat{\gamma}^{}_l \, \hat{O}_2  \, \right\rangle
    =  \left\langle   \, \hat{O}_1 \,
  \sum_{j}  \left(  \hat{\gamma}^{\dagger}_{j} \, \mathbb{M}_{jl}(\theta)
+  \hat{\gamma}^{}_{j} \, \mathbb{N}_{jl}(\theta) \right)
  \, e^{ \theta\hat{O} }  \, \hat{O}_2  \, \right\rangle
\end{equation}
and:
\begin{equation}
\label{Applendix:O1O2second}
    \left\langle   \, \hat{O}_1 e^{ \theta\hat{O} }  \, \, \hat{\gamma}^{\dagger}_l \, \hat{O}_2  \, \right\rangle
    =  \left\langle   \, \hat{O}_1 \,
  \sum_{j}  \left(  \hat{\gamma}^{\dagger}_{j} \, \mathbb{K}_{jl}(\theta)
+  \hat{\gamma}^{}_{j} \, \mathbb{L}_{jl}(\theta) \right)
  \, e^{ \theta\hat{O} }  \, \hat{O}_2  \, \right\rangle
\end{equation}
We need a bit of algebraic manipulations now. Let us multiply both sides of \eqref{Applendix:O1O2first} on the right by $\mathbb{N}^{-1} \mathbb{L}$:
\begin{equation}
 \begin{split}
     & \sum_{l} \left\langle   \, \hat{O}_1 e^{ \theta\hat{O} }  \, \, \hat{\gamma}^{}_l \, \hat{O}_2  \, \right\rangle \left( \mathbb{N}^{-1} \mathbb{L} \right)_{l,s} \\
     &
     = \sum_{l } \left\langle   \, \hat{O}_1 \,
  \sum_{j}  \left(  \hat{\gamma}^{\dagger}_{j} \, \mathbb{M}_{jl}(\theta) \right)
  \, e^{ \theta\hat{O} }  \, \hat{O}_2  \, \right\rangle \left( \mathbb{N}^{-1} \mathbb{L} \right)_{l,s} \\
  & + \left\langle   \, \hat{O}_1 \,
  \sum_{j}  \left(   \hat{\gamma}^{}_{j} \, \right)
  \, e^{ \theta\hat{O} }  \, \hat{O}_2  \, \right\rangle  \mathbb{L}_{j,s}
 \end{split}
\end{equation}
If we use this relation to replace the last term in \eqref{Applendix:O1O2second} we find the useful equality:
\begin{equation}
    \begin{split}
     &  \left\langle   \, \hat{O}_1 e^{ \theta\hat{O} }  \, \, \hat{\gamma}^{\dagger}_s \, \hat{O}_2  \, \right\rangle \\
  &  =   \left\langle   \, \hat{O}_1 \,
  \sum_{j}  \left(  \hat{\gamma}^{\dagger}_{j} \, \mathbb{K}_{js}(\theta) \right)
  \, e^{ \theta\hat{O} }  \, \hat{O}_2  \, \right\rangle \\
  & + \sum_{l} \left\langle   \, \hat{O}_1 e^{ \theta\hat{O} }  \, \, \hat{\gamma}^{}_l \, \hat{O}_2  \, \right\rangle \left( \mathbb{N}^{-1} \mathbb{L} \right)_{l,s} \\
  & - \sum_{l } \left\langle   \, \hat{O}_1 \,
  \sum_{j}  \left(  \hat{\gamma}^{\dagger}_{j} \, \mathbb{M}_{jl}(\theta) \right)
  \, e^{ \theta\hat{O} }  \, \hat{O}_2  \, \right\rangle \left( \mathbb{N}^{-1} \mathbb{L} \right)_{l,s}
    \end{split}
\end{equation}
Now, we consider the special case $\hat{O}_1 = 1$ and $\hat{O}_1 = \hat{\gamma}^{\dagger}_{s^{\prime}}$, and use the fact that all the creators acting on the vacuum on the right side give zero. Using the canonical anti-commutation relations we get:
\begin{equation}
\label{Appendix:gammadaggammadag}
 \left\langle   \,  e^{ \theta\hat{O} }  \, \, \hat{\gamma}^{\dagger}_s \,  \hat{\gamma}^{\dagger}_{s^{\prime}} \, \right\rangle 
 =  \left\langle   \,  e^{ \theta\hat{O} }  \, \right\rangle
 \left( \mathbb{N}^{-1} \mathbb{L} \right)_{s^{\prime},s}
\end{equation}
Now, it is interesting to find an expression for $ \left\langle   \,  e^{ \theta\hat{O} }  \, \right\rangle$. Let us compute:
\begin{equation}
   \begin{split}
   & \frac{d}{d\theta} \left\langle   \,  e^{ \theta\hat{O} }  \, \right\rangle =
   \left\langle   \,  e^{ \theta\hat{O} }  \, \hat{O} \, \right\rangle
   \end{split}
\end{equation}
As the operator $\hat{O}$ is acting on the vacuum, only the terms containing creation operators on the right give a non-zero contribution:
\begin{equation}
\label{Appendix:ddthetaeo}
\begin{split}
   &
 \frac{d}{d\theta} \left\langle   \,  e^{ \theta\hat{O} }  \, \right\rangle =
   \left\langle   \,  e^{ \theta\hat{O} }  \, \hat{O} \, \right\rangle \\
 & 
   = \frac{1}{2} \sum_{s, s^{\prime}} \left\langle   \,  e^{ \theta\hat{O} }  \, \mathcal{A}_{s,s^{\prime}} \,  \hat{\gamma}^{\dagger}_s \,  \hat{\gamma}^{\dagger}_{s^{\prime}}  \right\rangle 
   +  \frac{1}{2} \sum_{s, s^{\prime}} \left\langle   \,  e^{ \theta\hat{O} }  \, \left( -\mathcal{T}^{T} \right)_{s,s^{\prime}} \,  \hat{\gamma}^{}_s \,  \hat{\gamma}^{\dagger}_{s^{\prime}}  \right\rangle 
   \\
   & =  \frac{1}{2} \sum_{s, s^{\prime}} \mathcal{A}_{s,s^{\prime}} \,  \left\langle   \,  e^{ \theta\hat{O} }  \, \right\rangle
 \left( \mathbb{N}^{-1} \mathbb{L} \right)_{s^{\prime},s}
 - \frac{1}{2} Tr\left( \mathcal{T} \right)  \,  \left\langle   \,  e^{ \theta\hat{O} }  \, \right\rangle \\
 &
 = \left\langle   \,  e^{ \theta\hat{O} }  \, \right\rangle \, \frac{1}{2} \left( Tr \left( \mathcal{A} \mathbb{N}^{-1} \mathbb{L} \right) - Tr\left( \mathcal{T} \right)  \right)
    \end{split}
\end{equation}
where $\mathcal{A}$ and $\mathcal{T}$ are the $M \times M$ off-diagonal sub-blocks of the matrix:
\begin{equation}
 \mathcal{O} \buildrel{def}\over{=} 
 \left(
    \begin{array}{cc}
    \mathcal{T} & \mathcal{A} \\
    -\mathcal{A}^{\star} & - \mathcal{T}^{T}
    \end{array}
    \right)
\end{equation}
and we have used \eqref{Appendix:gammadaggammadag}.
Now, going back to \eqref{Appendix:Wtheta}, we can write:
\begin{equation}
    \mathcal{W(\theta + \varphi)} = e^{(\theta + \varphi) \hat{O}} = \mathcal{W(\theta)}\mathcal{W(\varphi)}
\end{equation}
We can differentiate both sides of this equality with respect to $\varphi$ in the point $\varphi = 0$:
\begin{equation}
    \mathcal{W}^{\prime}(\theta) =  \mathcal{W(\theta)}\mathcal{W}^{\prime}(0)
\end{equation}
Considering the lower right $M \times M$ sub-block of this matrix equation we find:
\begin{equation}
\label{Appendix:nprimetheta}
    \mathbb{N}^{\prime}(\theta) = \mathbb{L}(\theta)  \mathbb{M}^{\prime}(0)
    + \mathbb{N}(\theta)  \mathbb{N}^{\prime}(0)
\end{equation}
We can use the simple result:
\begin{equation}
   \mathcal{W}^{\prime}(0) = \frac{d}{d\varphi} e^{\varphi \mathcal{O}}|_{\varphi=0} = \mathcal{O}
\end{equation}
implying $\mathbb{M}^{\prime}(0) = \mathcal{A}$ and $\mathbb{N}^{\prime}(0) = - \mathcal{T}^{T}$. If we multiply both sides of \eqref{Appendix:nprimetheta} by $\mathbb{N}(\theta)^{-1}$ we find:
\begin{equation}
\label{Appendix:nm1nprimetheta}
   \mathbb{N}(\theta)^{-1} \mathbb{N}^{\prime}(\theta) = \mathbb{N}(\theta)^{-1}\mathbb{L}(\theta) \mathcal{A}
    -   \mathcal{T}^{T}
\end{equation}
Now, we can take the trace of both sides of the above relation. Using the cyclic property of the trace we obtain:
\begin{equation}
    Tr\left(\mathcal{A}\mathbb{N}(\theta)^{-1}\mathbb{L}(\theta)   \right)
    = Tr\left(\mathbb{N}(\theta)^{-1} \mathbb{N}^{\prime}(\theta)  \right) + Tr\left( \mathcal{T} \right) 
\end{equation}
If we use the fact that:
\begin{equation}
    Tr\left(\mathbb{N}(\theta)^{-1} \mathbb{N}^{\prime}(\theta)  \right) = 
    \frac{d}{d\theta} Tr\left( \log\mathbb{N}(\theta)  \right)
\end{equation}
and combine with \eqref{Appendix:ddthetaeo} we obtain the very interesting result:
\begin{equation}
    \frac{d}{d\theta} \left\langle   \,  e^{ \theta\hat{O} }  \, \right\rangle 
   = \left\langle   \,  e^{ \theta\hat{O} }   \, \right\rangle \, 
   \frac{1}{2} \left( \frac{d}{d\theta} Tr\left( \log\mathbb{N}(\theta)  \right)  \right)
\end{equation}
or:
\begin{equation}
    \frac{d}{d\theta} \log\left( \left\langle   \,  e^{ \theta\hat{O} }  \, \right\rangle \right)
   = 
   \frac{1}{2} \left( \frac{d}{d\theta} Tr\left( \log\mathbb{N}(\theta)  \right)  \right)
\end{equation}
Integrating both sides with respect to $\theta$ between $0$ and $\theta$, observing that $\mathbb{N}(\theta=0) = \mathbb{I}$, we find the very interesting result:
\begin{equation}
   \left\langle   \,  e^{ \theta\hat{O} }  \, \right\rangle
   = \exp\left(\frac{1}{2} Tr\left( \log\mathbb{N}(\theta)  \right)\right) 
\end{equation}
This completes the proof of \eqref{AppendixA:ExpectationExpO} if we set $\theta = 1$ and use the relation $Tr\left( \log\mathbb{N}  \right) = \log\left( \det\left( \mathbb{N} \right) \right)$.

\section{Factorization of exponentials of hermitian quadratic operators}
Using the notations in the previous Appendix, we will now prove the relation:
\begin{equation}
    \label{AppendixB:ExpOomegav}
     \exp(\hat{O})  \, | \, \omega_v \rangle
   = \sqrt{\det(\mathbb{N})} \, \exp\left( \frac{1}{2} \sum_{s,s^{\prime}} 
   \left( \mathbb{M} \mathbb{N}^{-1} \right)_{s,s^{\prime}} \,
   \hat{\gamma}^{\dagger}_{s} \hat{\gamma}^{\dagger}_{s^{\prime}} 
   \right) \, | \, \omega_v \rangle
\end{equation}

In order to prove this result we observe that the operators $\hat{\gamma}^{\dagger}\hat{\gamma}^{\dagger}$, $\hat{\gamma}^{\dagger} \hat{\gamma}^{}$ and $\hat{\gamma}^{} \hat{\gamma}^{}$ form an algebra which is closed under commutation, which implies that we are allowed to write:
\begin{equation}
    \begin{split}
    & \exp\left(  \frac{1}{2}  \left( \hat{\bf{\gamma}}^{\dagger} \,\, \hat{\bf{\gamma}}^{}   \right) \,
\left( \mathcal{O}  \right) \,
\left( 
\begin{array}{c}
\hat{\bf{\gamma}}^{} \\
\hat{\bf{\gamma}}^{\dagger}
\end{array} \right) \right) \\
    & = \alpha \, e^{ \frac{1}{2} \sum_{s,s^{\prime}} 
    \mathbb{Z}_{s,s^{\prime}} \,
   \hat{\gamma}^{\dagger}_{s} \hat{\gamma}^{\dagger}_{s^{\prime}} }\, e^{ \sum_{s,s^{\prime}} 
    \mathbb{X}_{s,s^{\prime}} \,
   \hat{\gamma}^{\dagger}_{s} \hat{\gamma}^{}_{s^{\prime}} } \, e^{ \sum_{s,s^{\prime}} 
    \mathbb{Y}_{s,s^{\prime}} \,
   \hat{\gamma}^{}_{s} \hat{\gamma}^{}_{s^{\prime}} }
    \end{split}
\end{equation}
where the coefficients are to be determined.
By taking the expectation value with respect to the vacuum and using \eqref{AppendixA:ExpectationExpO} we immediately find:
\begin{equation}
    \alpha = \left\langle   \,  e^{ \hat{O} }  \, \right\rangle = \sqrt{\det(\mathbb{N})}
\end{equation}
Applying both sides to the vacuum we find:
\begin{equation}
\label{AppendixB:eooualphaz}
    e^{ \hat{O} }  \,  | \, \omega_v \rangle = 
    \alpha \, e^{ \frac{1}{2} \sum_{s,s^{\prime}} 
    \mathbb{Z}_{s,s^{\prime}} \,
   \hat{\gamma}^{\dagger}_{s} \hat{\gamma}^{\dagger}_{s^{\prime}} }\, | \, \omega_v \rangle
\end{equation}
which implies that, for our purposes, we just need to find the coefficients $\mathbb{Z}_{s,s^{\prime}}$.
Returning to the relation \eqref{Applendix:O1O2first}, in the particular case $\hat{O}_1 = \hat{\gamma}^{}_{s^{\prime}}$ and $\hat{O}_2 = 1$:
\begin{equation}
    \left\langle  \hat{\gamma}^{}_{s^{\prime}} \,  e^{ \theta\hat{O} }  \, \, \hat{\gamma}^{}_l \,  \, \right\rangle
    =  \left\langle   \, \hat{\gamma}^{}_{s^{\prime}} \,
  \sum_{j}  \left(  \hat{\gamma}^{\dagger}_{j} \, \mathbb{M}_{jl}(\theta)
+  \hat{\gamma}^{}_{j} \, \mathbb{N}_{jl}(\theta) \right)
  \, e^{ \theta\hat{O} }  \,   \, \right\rangle
\end{equation}
The left-hand-side vanishes as a destructor is acting on the vacuum, so we have:
\begin{equation}
    \left\langle   \, \hat{\gamma}^{}_{s^{\prime}} \,
  \sum_{j}  \left(  \hat{\gamma}^{\dagger}_{j} \, \mathbb{M}_{jl}(\theta) \right)
  \, e^{ \theta\hat{O} }  \,   \, \right\rangle
  = - \left\langle   \, \hat{\gamma}^{}_{s^{\prime}} \,
  \sum_{j}  \left(  
  \hat{\gamma}^{}_{j} \, \mathbb{N}_{jl}(\theta) \right)
  \, e^{ \theta\hat{O} }  \,   \, \right\rangle
\end{equation}
We can multiply both sides by $\mathbb{N}^{-1}_{ls}(\theta)$ and sum over $l$, getting:
\begin{equation}
 \begin{split}
   &     \left\langle   \, \hat{\gamma}^{}_{s^{\prime}} \, 
  \hat{\gamma}^{}_{s} \,
  \, e^{ \theta\hat{O} }  \,   \, \right\rangle
  = -  \left\langle   \, \hat{\gamma}^{}_{s^{\prime}} \,
  \sum_{j}  \left(  \hat{\gamma}^{\dagger}_{j} \, \left(\mathbb{M}(\theta) \mathbb{N}^{-1}(\theta)\right)_{js} \right)
  \, e^{ \theta\hat{O} }  \,   \, \right\rangle \\
  & = - \left(\mathbb{M}(\theta) \mathbb{N}^{-1}(\theta)\right)_{s^{\prime}, s} \, \left\langle   \,
  \, e^{ \theta\hat{O} }  \,   \, \right\rangle
 \end{split}
\end{equation}
where in the last step we used canonical anti-commutation relations.
Incidentally, the formula above implies that the matrix $\mathbb{M}(\theta) \mathbb{N}^{-1}(\theta)$ must be anti-symmetric.
Now, we can combine with \eqref{AppendixB:eooualphaz}, remembering that $\alpha =\left\langle   \,
  \, e^{ \theta\hat{O} }  \,   \, \right\rangle $, to write:
\begin{equation}
  - \left(\mathbb{M}(\theta) \mathbb{N}^{-1}(\theta)\right)_{s^{\prime}, s}
  = \frac{ \left\langle   \, \hat{\gamma}^{}_{s^{\prime}} \, 
  \hat{\gamma}^{}_{s} \,
  \, e^{ \theta\hat{O} }  \,   \, \right\rangle}{\left\langle   \,
  \, e^{ \theta\hat{O} }  \,   \, \right\rangle}
  = \left\langle   \, \hat{\gamma}^{}_{s^{\prime}} \, 
  \hat{\gamma}^{}_{s} \,
  \, e^{ \frac{1}{2} \sum 
    \mathbb{Z}_{l,l^{\prime}} \,
   \hat{\gamma}^{\dagger}_{l} \hat{\gamma}^{\dagger}_{l^{\prime}} }\,  \,   \, \right\rangle
\end{equation}
If we expand the exponential in the last term and use the canonical anti-commutation relations, we easily conclude that only the term with two destructors will give a non-zero result. In particular, we get:
\begin{equation}
    \begin{split}
      &   \left\langle   \, \hat{\gamma}^{}_{s^{\prime}} \, 
  \hat{\gamma}^{}_{s} \,
  \, e^{ \frac{1}{2} \sum 
    \mathbb{Z}_{l,l^{\prime}} \,
   \hat{\gamma}^{\dagger}_{l} \hat{\gamma}^{\dagger}_{l^{\prime}} }\,  \,   \, \right\rangle = \left\langle   \, \hat{\gamma}^{}_{s^{\prime}} \, 
  \hat{\gamma}^{}_{s} \,
  \,  \frac{1}{2} \sum_{l,l^{\prime}}
    \mathbb{Z}_{l,l^{\prime}} \,
   \hat{\gamma}^{\dagger}_{l} \hat{\gamma}^{\dagger}_{l^{\prime}} \,  \,   \, \right\rangle \\
   & = \frac{1}{2} \sum_{l,l^{\prime}} \mathbb{Z}_{l,l^{\prime}} \, \left( \delta_{s,l} \delta_{s^{\prime}, l^{\prime}}
    - \delta_{s^{\prime},l} \delta_{s, l^{\prime}} \right) = \frac{1}{2} \left( \mathbb{Z}_{s,s^{\prime}} - \mathbb{Z}_{s^{\prime},s} \right)
    \end{split}
\end{equation}
We observe that in an expansion of the form $\sum_{l,l^{\prime}}
    \mathbb{Z}_{l,l^{\prime}} \,
   \hat{\gamma}^{\dagger}_{l} \hat{\gamma}^{\dagger}_{l^{\prime}} 
   = \frac{1}{2} \left(  \mathbb{Z}_{l,l^{\prime}} \,
   \hat{\gamma}^{\dagger}_{l} \hat{\gamma}^{\dagger}_{l^{\prime}}
   -  \mathbb{Z}_{l,l^{\prime}} \,
   \hat{\gamma}^{\dagger}_{l^{\prime}} \hat{\gamma}^{\dagger}_{l} \right)$
we can always choose $\mathbb{Z}$ to be anti-symmetric. We thus find $\mathbb{Z}_{s^{\prime},s} = \left(\mathbb{M}(\theta) \mathbb{N}^{-1}(\theta)\right)_{s^{\prime}, s}$, which completes the proof of \eqref{AppendixB:ExpOomegav}.

\providecommand{\noopsort}[1]{}\providecommand{\singleletter}[1]{#1}%

\end{document}